\documentclass[aps,superscriptaddress,amsmath,amssymb,pra,twocolumn,10pt,floatfix]{revtex4}
\usepackage[dvips]{graphicx}
\usepackage{epsfig}
\usepackage{amssymb,amsfonts,amsmath,bm,rotating}

\begin{document}

\title{Symmetry analysis of crystalline spin textures in dipolar spinor condensates}
\author{R. W. Cherng}
\author{E. Demler}
\affiliation{Physics Department, Harvard University, Cambridge, MA 02138}
\date{\today}

\begin{abstract}
We study periodic crystalline spin textures in spinor condensates
with dipolar interactions via a systematic symmetry analysis of the
low-energy effective theory.
By considering symmetry operations which combine 
real and spin space operations, we classify 
symmetry groups consistent with non-trivial experimental
and theoretical constraints.
Minimizing the energy within each symmetry class
allows us to explore possible ground states.
\end{abstract}
\maketitle

\section{Introduction}

Experiments in ultracold atomic gases have provided direct and striking
evidence for the theory of Bose-Einstein condensation.
Typically, the combination of low temperatures and strong magnetic fields 
freezes out the internal level structure leaving only the 
density and phase as relevant degrees of freedom.  
However, recent experimental advancements for multicomponent condensates
include optical dipole traps used for preparation \cite{mit-98a} and
phase-contrast imaging used for detection \cite{berkeley-05} in $S=1$ $^{87}$Rb.

The magnetization, a vector quantity sensitive to both populations
and coherences between hyperfine levels, can be directly  imaged in these systems.
This has allowed the Berkeley group to observe evidence for
spontaneous formation of crystalline magnetic order \cite{dsk-08,dsk-09}.
When an initially incoherent gas is cooled below the critical
temperature, a crystalline lattice of spin domains emerges spontaneously
at sufficiently long times. 

Several theoretical studies have stressed the role of 
the effective dipolar interactions \cite{pfau-02,meystre-01,pu-06,ueda-06} 
strongly modified by magnetic field induced rapid Larmor precession
and reduced dimensionality \cite{cherng-09,ueda-09,ho-09}.
This can drive dynamical instabilities in a uniform condensate with
characteristic unstable modes at wavevectors in a pattern consistent
with observed magnetization correlations \cite{cherng-09,ueda-09}.
Numerical simulation of the full multicomponent mean-field dynamics
also suggests long-lived spin textures \cite{ueda-09,ho-09}.

In this paper, we take an alternative approach and 
focus directly on the low-energy degrees of freedom.  In a companion paper 
\cite{cherng-10a},
we derived a non-linear sigma model
describing the dynamics of the magnetization.
Due to coupling of the magnetization and superfluid velocity,
this effective theory includes a long ranged interaction between
skyrmions, topological objects familiar from the theory of
ferromagnets \cite{rajamaran-82}.  
For spinor condensates however, non-zero skyrmion density is 
directly associated with persistent, circulating superfluid currents.

\begin{figure}
\begin{center}
\includegraphics[width=3in]{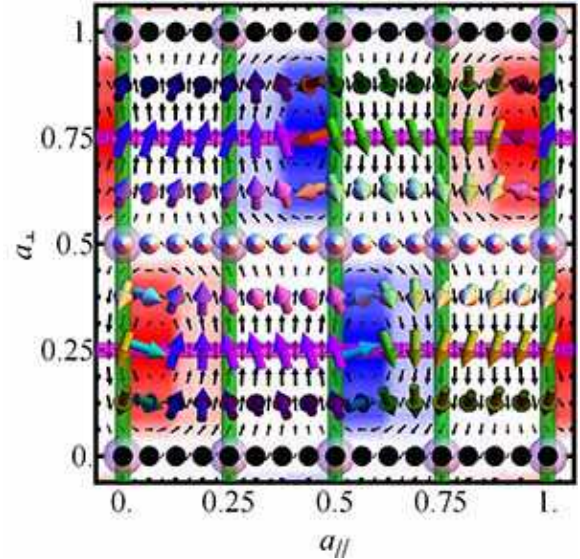}
\end{center}
\caption[Unit cell for minimal energy crystalline spin texture]{
Unit cell for the minimal energy crystalline spin texture.
The magnetice field $\hat{B}$ is along the horizontal axis and
lattice constants are $a_\parallel=90$ $\mu$m, $a_\perp=42$ $\mu$m.
Green lines indicating 
glide reflection lines, purple lines
indicating mirror lines, and
white spheres indicating rotation
centers describe symmetry operations.
See Fig. \ref{chap:symmetry:fig:symmetry} for a depiction 
of symmetry operations.
Red (blue) background indicates positive (negative) skyrmion 
density $q$, black 2D arrows the superfluid velocity $\mathbf{v}$, 
and shaded 3D arrows the magnetization $\hat{n}$ with
white (black) along $+\hat{B}$ ($-\hat{B}$) and hue indicating
orientation perpendicular to $\hat{B}$.
}
\label{chap:symmetry:fig:minimal_x}
\end{figure}

Our approach to the daunting task of exploring the
space of possible ground states is via a systematic symmetry
analysis which breaks up this space into distinct symmetry classes.
Each of these classes is characterized by invariance under a symmetry group
containing combined real space and spin space operations.
Litvin and Opechowski called these groups
the spin groups \cite{litvin-74}, a notation we will also use throughout
this paper \footnote{Here spin groups do not refer to the double cover
of the orthogonal group that arises in the theory of Lie groups}.
The focus of their paper on the study of magnetically ordered crystals.
In such systems, the spin degrees of freedom are localized at discrete atoms.
For spinor condensates,
the spin-dependent contact interaction which
determines the spin healing length is larger than the dipolar interaction strength 
which determines the size of individual spin domains.  Thus we are primarily
interested in smooth spin textures and will use spin groups in a novel
manner to classify them into distinct symmetry classes.

The power of using spin groups becomes apparent when we consider the 
non-trivial constraints that spin textures must satisfy.  Theoretical
constraints such as a non-vanishing magnetization must be satisfied
in order for the low-energy effective theory to be valid.
In addition, there are constraints coming from experimental observations
such as a vanishing net magnetization.
Only a relatively small number of spin groups are compatible with all
of these theoretical and experimental constaints and identifying them
allows us to significantly narrow the space of possible spin textures.

After identifying the allowed symmetry classes, we then minimize the 
energy for spin textures within class.  This allows us to obtain
crystalline spin textures as in Fig. \ref{chap:symmetry:fig:minimal_x}
which we find to have the lowest energy for current experimental parameters.
These numerical solutions including dipolar interactions
are qualitatively similar to the complementary analytical solutions
studied in the companion paper \cite{cherng-10a}.
These latter solutions in the absence of dipolar interactions
describe periodic configurations of topological objects called skyrmions.
The combined results
provide a consistent physical picture of the role of dipolar interactions 
in stabilizing non-trivial crystalline spin textures.  In particular,
such spin textures can be viewed as a lattice of smooth topological
objects carrying persistent superfluid currents.

\section{Hamiltonian}

Here we briefly review the non-linear sigma model describing
dipolar spinor condensates derived in the companion paper \cite{cherng-10a}.
We consider $S=1$ dipolar spinor condensates in a quasi-two-dimensional
geometry.  Below the scale of spin-independent and
spin-dependent contact interactions, the local density is fixed 
and the magnetization is maximally polarized.
Competition between the quadratic Zeeman shift and dipolar interactions
determines the formation of spin textures.
The following non-linear sigma model describes the effective theory
\small
\begin{align}
\nonumber
\mathcal{L}&=
\rho_{2D}\left[-\int dt d^2x \mathcal{A}(\hat{n})\cdot\partial_t\hat{n}
-\int dt \mathcal{H}_{KE}-\int dt \mathcal{H}_{S}\right]\\
\nonumber
\mathcal{H}_{KE}&=\frac{1}{4m}\int d^2x 
(\nabla\hat{n})^2
+\frac{1}{2m}\int d^2x d^2y q(x)G(x-y)q(y)\\
\mathcal{H}_{S}&=\int d^2x d^2y \hat{n}^i(x)h^{ij}(x-y)\hat{n}^j(y)
\label{chap:symmetry:eq:lagrangian}
\end{align}
\normalsize
where the magnetization $\hat{n}$ is a three component real unit vector,
$\mathcal{A}(\hat{n})$ is the unit monopole vector potential,
$\mathcal{H}_{KE}$ gives kinetic energy contributions,
$\mathcal{H}_{S}$ gives spin-dependent interactions,
and $\rho_{2D}$ is the two-dimensional density.

The first term in $\mathcal{H}_{KE}$ is the spin stiffness
while the second term comes form the superfluid kinetic energy.  
Non-uniform textures in $\hat{n}$ arise in part due to phase gradients of the 
underlying condensate wavefunction.
The resulting coupling of $\hat{n}$ to the superfluid velocity $\mathbf{v}$
fixes the vorticity $\epsilon_{\mu\nu}\nabla_\mu\mathbf{v}_\nu=q$
to the skyrmion density
\begin{align}
\label{chap:symmetry:eq:q}
q=\epsilon_{\mu\nu} \hat{n}\cdot \nabla_\mu \hat{n}\times \nabla_\nu\hat{n}
\end{align}
whose integral is a quantized topological invariant.
The superfluid kinetic energy becomes a logarithmic $G(x-y)$ vortex
interaction for $q$ where $-\nabla^2G(x-y)=\delta(x-y)$. 
Physically, gapless superfluid phase fluctuations generate the long-wavelength
divergence of $G(x-y)$.

For $\mathcal{H}_S$, the momentum space interaction tensor is \cite{cherng-09}
\small
\begin{align}
\label{chap:symmetry:eq:dipolar_effective}
\nonumber
h^{ij}(k)=&\tilde{Q}\left(\delta^{ij}+\hat{B}^i\hat{B}^j\right)
-\tilde{g}_d\left[\frac{3h(kd_n)-1}{2}\right]
\left[\delta^{ij}-3\hat{B}^i\hat{B}^j\right],\\
\nonumber
h(\vec{k})=&[\hat{B}\cdot \hat{k}]^2 w(k)+
[\hat{B}\cdot \hat{n}]^2[1-w(k)],\\
w(x)=&2x\int_0^\infty dz e^{-(z^2+2zx)}
\end{align}
\normalsize
where $\hat{B}$ is a unit vector along the magnetic field,
$d_n$ is the thickness of the condensate along the normal direction
which we assume to have a Gaussian form,
$\tilde{Q}=Q/2$ with $Q$ the quadratic Zeeman shift, 
$\tilde{g}_d=4\pi g_d n_{3D}C/3$ with
$g_d$ the dipolar interaction stength,
$n_{3D}$ the peak three-dimensional density, and $C=1/\sqrt{2}$ is
determined by normalization.
For current experimenets \cite{dsk-08,dsk-09}, $g_d n_{3D}=0.8$ Hz, $Q=1.5$ Hz, 
and $\hat{B}$ is in the plane.
For large quadratic Zeeman shifts, all atoms go into the $m_z=0$ state.  This
limits our analysis to the small $q$ regime.

\section{\label{chap:symmetry:sec:constraints}Spin texture constraints}

Minimizing the above Hamiltonian is difficult due to a number of non-trivial
constraints on possible spin textures.  We first consider 
\textit{fundamental constraints} coming from theoretical considerations
for a valid low-energy effective theory.

The first is given by
(a) zero net skyrmion charge $\int d^2x q=0$. This
arises due to the long-wavelength divergence of the skyrmion interaction.
Recall the skyrmion density acts as a source for superfluid vorticity.
The logarithmic interaction between vortices in two dimensions implies
that only net neutral configurations of skyrmions have finite energy.

The second is given by 
(b) maximally polarized magnetization $|\hat{n}|=1$.
Recall the non-linear sigma model derived in the companion paper \cite{cherng-10a}
is valid in the regime where the spin-dependent contact interaction
is larger than the dipolar interaction and quadratic Zeeman shift.
In this regime, the spin-dependent contact interaction favors a local 
magnetization that is maximally polarized while the dipolar and Zeeman terms
determine the local orientation of the spin texture.

The third is given by
(c) explicit symmetry breaking of spin rotational invariance by $\hat{B}$
and the dipolar interaction.  In the absence of the dipolar interaction
and applied magnetic field, the system is invariant under independent
spin-space and real-space rotations/refelctions.  The external field along $\hat{B}$ 
explicitly breaks the spin space symmetry down to rotations/reflections
that fix $\hat{B}$ in spin space.
For the bare dipolar interaction, the spin-orbit coupling implies
that only combined spin-space and real-space rotations remain a symmetry.
However, the effective dipolar interaction is averaged by rapid Larmor 
precession about the axis of the applied magnetic field $\hat{B}$.  The combined
effect of the effective dipolar interaction and external field $\hat{B}$ is that
independent and arbitrary real-space rotations and spin-space rotations are
explicitly broken down to independent real-space rotations and spin-space 
rotations that fix $\hat{B}$. 

Next we consider \textit{phenomenological constraints} 
coming from properties of the Berkeley group's experimentally observed spin 
textures.
We focus on the spin textures prepared by cooling from the incoherent
high-temperature equilibrium state with each hyperfine level having equal 
initial populations
\cite{dsk-09}.

The fourth constraint is given by
(d) periodic crystalline order with a rectangular lattice.
Direct real-space imaging of the spin textures shows evidence for a lattice
of spin domains.  The resulting spin correlation function shows strong peaks
in a characteristic cross-like pattern suggestive of a rectangular unit cell.

The fifth is 
(e) spin textures are not easy axis nor easy plane but cover spin space.
All three components of the magnetization can be imaged within the same sample
and shows evidence that the spin texture is not confined to vary only 
along a single axis or a single plane.

The sixth and final is
(f) zero net magnetization $\int d^2x \hat{n}=0$.
The distribution of the magnetization vector shows modulations are centered
about zero and yield no net magnetization.
We note that the Berkeley group has considered spin textures prepared
from a non-equilibrium state with imbalanced initial populations.  The
resulting spin textures carry a net magnetization.  Although we do not
consider such spin textures directly, they can be studied within the
same symmetry analysis framework we describe below.

\section{\label{chap:symmetry:sec:sg}Space groups and spin groups }
Having considered the spin texture constraints, we now 
describe the structure of space groups and their generalization to 
spin groups in two dimensions.  Originally
developed in crystallography, we will use them to study
smooth spin textures.  In particular, we will show in the next section
that there are only a small number of compatible spin groups
consistent with the above constraints.
For a brief overview of group theory and representation theory,
see Appendix \ref{chap:symmetry:app:reps}.

Crystals of featureless atoms with no internal degrees of freedom
can be classified by space groups.  For more details on space groups,
see Ref. \cite{cracknell-72,kim-99}.
It is instructive to consider space groups as subgroups
of $E(2)$, the two-dimensional Euclidean group of real-space
translations and rotations/reflections.  We first describe the
elements of $E(2)$.
The real-space translations are given by a two-component vector $t$ while the 
rotations/reflections are given by a two-by-two orthogonal matrix $M$.
The resulting group element $(M,t)$ acts on a two-component position
$x$ as
\begin{align}
x_\mu\rightarrow M_{\mu\nu}x_\nu+t_\mu
\end{align}
which shows that the product of two elements in $E(2)$ is given by
\begin{align}
(M',t')(M,t)=(M'M,M't+t')
\end{align}
and notice that the real-space rotation has a non-trivial
action on the real-space translation.
Crystals do not have continuous translation and continuous rotation
symmetries of $E(2)$.  They describe spontaneous breaking of $E(2)$
down to a discrete set of translations and rotations/reflections
called a space group.

First consider groups formed from discrete translations.
This forms the Bravais lattice and 
can be written in terms of the generators $t_1$, $t_2$ as $t=c t_1+d t_2$ where
$c$ and $d$ are integers and $t_1$, $t_2$ are two component vectors.  
In two dimensions, there are five distinct
Bravais lattices: oblique, rectangular, centered, square, and hexagonal.

Next consider groups formed from discrete rotations/reflections.
This forms the point group and can
be written in terms of the generators $r$ and $s$ for rotations and reflections
as $M=r^as^b$ where $a$ and $b$ are integers and $r$, $s$ are two-by-two
orthogonal matrices.
In two dimensions, there are two classes of point groups: cyclic groups $C_n$ 
of $2\pi/n$ rotations and dihedral groups $D_n$ of $2\pi/n$ rotations and 
reflections.
For $C_n$ the generators satisfy $r^n=s=\mathbf{1}$ while
for $D_n$ the generators satisfy $r^n=s^2=(rs)^2=\mathbf{1}$ where
$\mathbf{1}$ is the identity element.
The order $n$ of $C_n$ and $D_n$ are restricted to $n=1,2,3,4,6$.
A more detailed discussion of point groups in two dimensions is given in
Appendix \ref{chap:symmetry:app:pg}.

Notice the Bravais lattice and the point group contain only pure
translations and pure rotations/reflections, respectively.
Since rotations/reflections can act non-trivially on translations,
a space group specifies additional information on how
to combine the Bravais lattice and point 
group.  Formally, the Bravais lattice $T$ is a normal subgroup of the 
space group $SG$ and the proint group $PG$ is the quotient group $PG=SG/T$.
In particular,
the space group itself can contain non-trivial combinations of translation
and rotation/reflection operations.  When this is the case, the space group
is called non-symmorphic, otherwise it is symmorphic.  Viewing the generators
$t_1$, $t_2$ of the Bravias lattice as elements $T_1$, $T_2$ of the 
space group we can write
\begin{align}
\label{chap:symmetry:eq:translations}
T_1&=\left(\mathbf{1},t_1\right),& T_2=\left(\mathbf{1},t_2\right)
\end{align}
where $t_i$ is a two-component vector and $\mathbf{1}$ is the
$2\times2$ identity matrix.
Viewing the generators of $r$, $s$ of the point group as elements $R$, $S$ 
of the space group we can write
\begin{align}
\label{chap:symmetry:eq:rotations1}
R&=\left(s(\theta_R),n^R_1t_1+n^R_2t_2\right),&
S&=\left(r(\theta_S),n^S_1t_1+n^S_2t_2\right)
\end{align}
where the $2\times2$ rotation and reflection matrices are given by
\begin{align}
\label{chap:symmetry:eq:rotations2}
\nonumber
r(\theta_R)&=
\begin{bmatrix}
\cos(\theta)&-\sin(\theta)\\
\sin(\theta)&\cos(\theta)
\end{bmatrix},\\
s(\theta_S)&=
\begin{bmatrix}
-\cos(2\theta)&-\sin(2\theta)\\
-\sin(2\theta)&\cos(2\theta)
\end{bmatrix}
\end{align}
respectively.  Then the most general element of the space group
is written as
\begin{align}
(M,t)=R^aS^bT_1^cT_2^d
\label{chap:symmetry:eq:space_group}
\end{align}
where $a$,$b$,$c$,$d$ are integers.  There are 17 distinct
space groups and the corresponding parameters are adapted from
Ref. \cite{litvin-08} and given in
Table \ref{chap:symmetry:tab:space_group}.

\begin{table*}
\begin{center}
\begin{tabular}{llllllllllll}
$SG$&Type&
Lat.& $t_1$& $t_2$& 
$PG$& 
$\theta_R$&$n^R_1$&$n^R_2$& 
$\theta_S$&$n^S_1$&$n^S_2$\\
\hline
$p1$&sym&
Oblique&$(a\cos(\gamma),a\sin(\gamma))$&$(0,b)$&
$C_1$& 0& $0$&$0$&---&---&---\\

$p211$&sym&
Oblique&$(a\cos(\gamma),a\sin(\gamma))$&$(0,b)$&
$C_2$& $\pi$& $0$&$0$&---&---&---\\

$p1m1$&sym&
Rectangular&$(a,0)$&$(0,b)$&
$D_1$& 0& $0$&$0$& 0&$0$&$0$\\

$p1g1$&non&
Rectangular&$(a,0)$&$(0,b)$&
$D_1$& 0& $0$&$0$& 0& $0$&$1/2$\\

$c1m1$&sym&
Centered&$(a/2,b/2)$&$(0,b)$&
$D_1$& 0& $0$&$0$& 0&$0$&$0$\\

$p2mm$&sym&
Rectangular&$(a,0)$&$(0,b)$&
$D_2$& $\pi$& $0$&$0$& 0&$0$&$0$\\

$p2mg$&non&
Rectangular&$(a,0)$&$(0,b)$&
$D_2$& $\pi$& $0$&$0$& 0& $1/2$&$0$\\

$p2gg$&non&
Rectangular&$(a,0)$&$(0,b)$&
$D_2$& $\pi$& $0$&$0$& 0& $1/2$&$1/2$\\

$c2mm$&sym&
Centered&$(a/2,b/2)$&$(0,b)$&
$D_2$& $\pi$& $0$&$0$& 0&$0$&$0$\\

$p4$&sym&
Square&$(a,0)$&$(0,a)$&
$C_4$& $\pi/2$& $0$&$0$&---&---&---\\

$p4mm$&sym&
Square&$(a,0)$&$(0,a)$&
$D_4$& $\pi/2$& $0$&$0$& 0&$0$&$0$\\

$p4gm$&non&
Square&$(a,0)$&$(0,a)$&
$D_4$& $\pi/2$& $0$&$0$& 0& $1/2$&$1/2$\\

$p3$&sym&
Hexagonal&$(a\sqrt{3}/2,-a/2)$&$(0,a)$&
$C_3$& $2\pi/3$& $0$&$0$&---&---&---\\

$p3m1$&sym&
Hexagonal&$(a\sqrt{3}/2,-a/2)$&$(0,a)$&
$D_3$& $2\pi/3$& $0$&$0$& $-\pi/6$&$0$&$0$\\

$p31m$&sym&
Hexagonal&$(a\sqrt{3}/2,-a/2)$&$(0,a)$&
$D_3$& $2\pi/3$& $0$&$0$& 0&$0$&$0$\\

$p6$&sym&
Hexagonal&$(a\sqrt{3}/2,-a/2)$&$(0,a)$&
$C_6$& $\pi/3$& $0$&$0$&---&---&---\\

$p6mm$&sym&
Hexagonal&$(a\sqrt{3}/2,-a/2)$&$(0,a)$&
$D_6$& $\pi/3$& $0$&$0$& $-\pi/6$&$0$&$0$\\
\end{tabular}
\end{center}
\caption[Two-dimensional space groups]{The seventeen
two-dimensional 
space groups ($SG$) have elements of the
form $(M,t)=R^aS^bT_1^cT_2^d$.  Here $a$, $b$, $c$, $d$ are integers.
Each space group is one of two types (Type)
symmorphic (sym.) or non-symmorphic (non).
The normal subgroup of translations $T$ is one of four
Bravais lattice types (Lat.) with the generators
$T_1$, $T_2$.
The quotient group $SG/T$ is the point group ($PG$)
and has generators $R$, $S$. 
The parameters $t_1$, $t_2$ specify the generators $T_1$, $T_2$
through Eq. \ref{chap:symmetry:eq:translations}.
The parameters $\theta_{R,S}$, $\theta_{R,S}$, $n^{R,S}_{1,2}$
specify the generators $R$, $S$ through Eqs.
\ref{chap:symmetry:eq:rotations1}, \ref{chap:symmetry:eq:rotations2}.
Adapted from Ref. \cite{litvin-08}.}
\label{chap:symmetry:tab:space_group}
\end{table*}

Litvin and Opechowski \cite{litvin-74} considered the classification
of magnetically ordered crystals 
of atoms with internal spin degrees of freedom via spin groups.  These
groups are generalizations of space groups with combined real space 
translations, real-space rotations/reflections, as well as
spin-space rotations/reflections.  Here we consider how they can be explicitly
constructed from the representation theory of space groups more suitable
for calculations.  Litvin and Opechowski consider
a more implicit classification of spin groups which we show is equivalent
in Appendix \ref{chap:symmetry:app:sg}.

It is instructive to consider spin groups as subgroups of the direct
product $E(2)\otimes O(3)$, where $E(2)$ is the two-dimensional
Euclidean group of real-space
translations and rotations/reflections and $O(3)$ is the
three-dimensional orthogonal group of spin-space rotations/reflections.
Recall the real-space translations are given by a two-component vector $t$ 
while the rotations/reflections are given by a two-by-two orthogonal matrix $M$.
In addition the spin-space rotations/reflections are given by a three-by-three
orthogonal matrix $O$.
The resulting group element $(M,t,O)$ acts on a three-component spin $\hat{n}(x)$
that is a function of a two-component position $x$ as
\begin{align}
\label{chap:symmetry:eq:sg_action}
\hat{n}^i(x_\mu)&\rightarrow O^{ij}\hat{n}^j(M_{\mu\nu}x_\nu+t_\mu)
\end{align}
which shows that the product of two elements in $E(2)\otimes O(3)$ is given
by
\begin{align}
(M',t',O')(M,t,O)=(M'M,M't+t',O'O)
\label{chap:symmetry:eq:space_group_product}
\end{align}
and notice that while the real-space rotation has a non-trivial
action on the real-space translation, the real-space and spin-space operations
do not act on each other.
Magnetically ordered crystals do not have continuous real-space translation,
real-space rotation, and continuous spin-space rotations 
symmetries of $E(2)\otimes O(3)$.  They describe spontaneous breaking of 
$E(2)\otimes O(3)$
down to a discrete set of real-space translations, real-space rotations/reflections,
and spin-space rotations/reflections called a spin group.

To construct spin groups, start by choosing a space group $SG$ 
giving the real-space operations.
Now choose a three-dimensional orthogonal representation $\phi$ of the
space group $SG$.
This is a function from $SG$
to three-dimensional orthogonal matrices satisfying the homomorphism
condition
\begin{align}
\phi(M',t')\phi(M,t)=\phi(M'M,M't+t')
\label{chap:symmetry:eq:homomorphism}
\end{align}
For this representation $\phi$, choose a group of three-dimensional
orthogonal matrices $N$ that satisfies
\begin{align}
\phi(M,t)^{-1}N\phi(M,t)=N
\end{align}
consisting of three-by-three orthogonal matrices that are left fixed
under conjugation by $\phi(M,t)$ for all elements $(M,t)$ of the 
the space group $SG$.
The resulting spin group has elements of the form
\begin{align}
\label{chap:symmetry:eq:spin_group}
(M,t,O)=(M,t,n\phi(M,t))
\end{align}
where $(M,t)$ are the elements of a space group $SG$,
$\phi$ is a representation of $SG$, and $n$ is an element of $N$.
The most general space group element is of the form in 
Eq. \ref{chap:symmetry:eq:space_group}.
Using the space group product of Eq. \ref{chap:symmetry:eq:space_group}
and homomorphism condition Eq. \ref{chap:symmetry:eq:homomorphism},
we see that 
\begin{align}
\label{eq:symmetry:eq:hom_gen}
\phi(R^aS^bT_1^cT_2^d)=\phi(R)^a\phi(S)^b\phi(T_1)^c\phi(T_2)^d
\end{align}
meaning we only need to specify the values of the representation
on the space group generators.

\section{\label{chap:symmetry:sec:compatible}Compatible spin groups}

Before discussing how to impose the constraints of Sec. 
\ref{chap:symmetry:sec:constraints}, we first discuss the physical
interpretation of the structure of spin groups.  Recall
that a spin group is given by a choice of space group $SG$ with elements $(M,t)$, 
three-dimensional orthogonal representation $\phi$, and a
choice of three-dimensional orthogonal matrices $N$ that commute
as a set with each $\phi(M,t)$.

First consider the group $N$.  From Eq. \ref{chap:symmetry:eq:spin_group},
we see that by taking $M=\mathbf{1}$ with $\mathbf{1}$ the $2\times 2$
identity matrix and $t=0$, the spin group contains
the elements $(\mathbf{1},0,n)$ where $n$ is an element of $N$.  The physical
interpretation is that $N$ describes global spin-space symmetries that
do not act on spatial degrees of freedom.  For example, a uniform magnetization
is described by $N$ containing rotations and reflections that leave the magnetization
fixed.

Next consider the group given by the kernel $\text{ker}(\phi)$ of the 
representation.
This consists of elements $(M',t')$ that satisfy $\phi(M',t')=\mathbf{1}$
with $\mathbf{1}$ the $3\times 3$ identity matrix.  These elements
form a space group $SG'$ that is a subgroup of $SG$.  From 
Eq. \ref{chap:symmetry:eq:spin_group}, we see that the spin group
contains the elements $(M',t',\mathbf{1})$.  The physical interpretation
is that $SG'$ describes global real-space symmetries that do not
act on spin degrees of freedom.  The distinction between $SG$ and $SG'$
is that $SG$ describes the symmetries of the \textit{crystallographic unit cell}
while $SG'$ describes the symmetries of the \textit{magnetic unit cell}.

Consider a square lattice with lattice constant $a$ and one spinful atom 
per unit cell and anti-ferromagnetic order.
The crystallographic unit cell is generated by the vectors $(0,a)$ and $(a,0)$
and contains one spinful atom.  This is the unit cell ignoring spin
and described by a space group $SG$.
The magnetic unit cell is generated by the vectors $(+a,-a)$ and $(-a,+a)$
and contains two spinful atoms.
This is the unit cell taking into account spin and described by a space
group $SG'$ that is a subgroup of $SG$.

From now on, we focus on applications of spin groups to classify
smooth spin textures.
In order to understand which spin groups are compatible with the
constraints discussed earlier, we consider how these symmetry operations
act on the magnetization vector $\hat{n}$ and skyrmion density $q$.
In real space, an element $(M,t,O)$ of a spin group acts as
\begin{align}
\nonumber
\hat{n}^i(x_\mu)&\rightarrow O^{ij}\hat{n}^j(M_{\mu\nu}x_\nu+t_\mu),\\
q(x_\mu)&\rightarrow \det[O]\det[M]q(M_{\mu\nu}x_\nu+t_\mu)
\label{chap:symmetry:eq:action_x}
\end{align}
where we have used Eq. \ref{chap:symmetry:eq:sg_action} for the action
on $\hat{n}$ which along with Eq. \ref{chap:symmetry:eq:q} allows
us to deduce the action on $q$.
In momentum space, the action is
\begin{align}
\nonumber
\hat{n}^i(k_\mu)&\rightarrow\exp(ik_\mu M^{-1}_{\mu\nu}t_\nu)
O^{ij}\hat{n}^j(M_{\mu\nu}k_\nu),\\
q(k_\mu)&\rightarrow\exp(ik_\mu M^{-1}_{\mu\nu}t_\nu)
\det[O]\det[M]q(M_{\mu\nu}k_\nu)
\label{chap:symmetry:eq:action_k}
\end{align}
which follow directly from the Fourier transform.

\begin{figure}
\begin{center}
\begin{tabular}{cccc}
\begin{sideways}
\textbf{(a)} Reflection
\end{sideways}&
\includegraphics[width=1.5in]{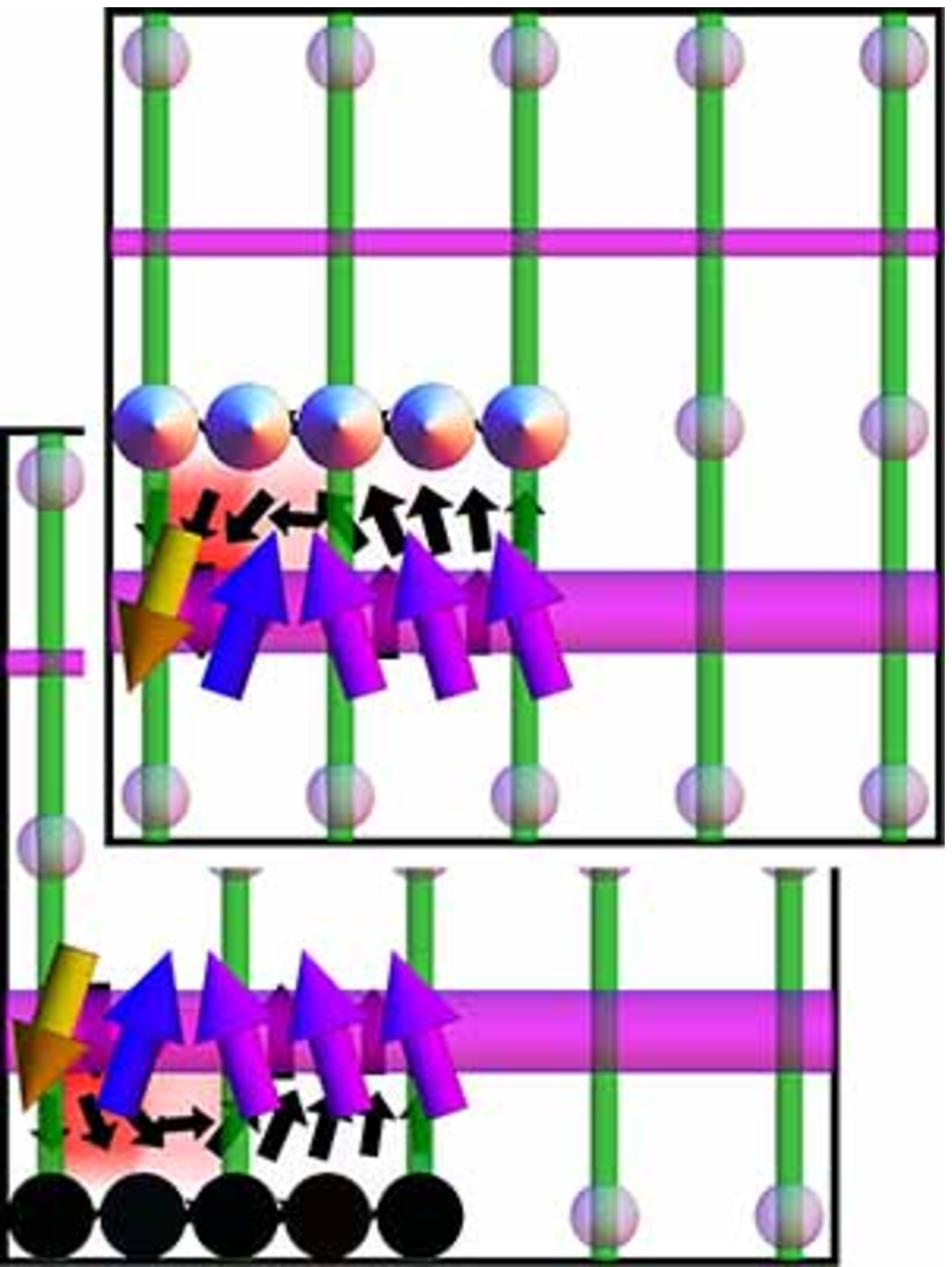}&
\begin{sideways}\textbf{(b)}
Glide reflection 
\end{sideways}&
\includegraphics[width=1.5in]{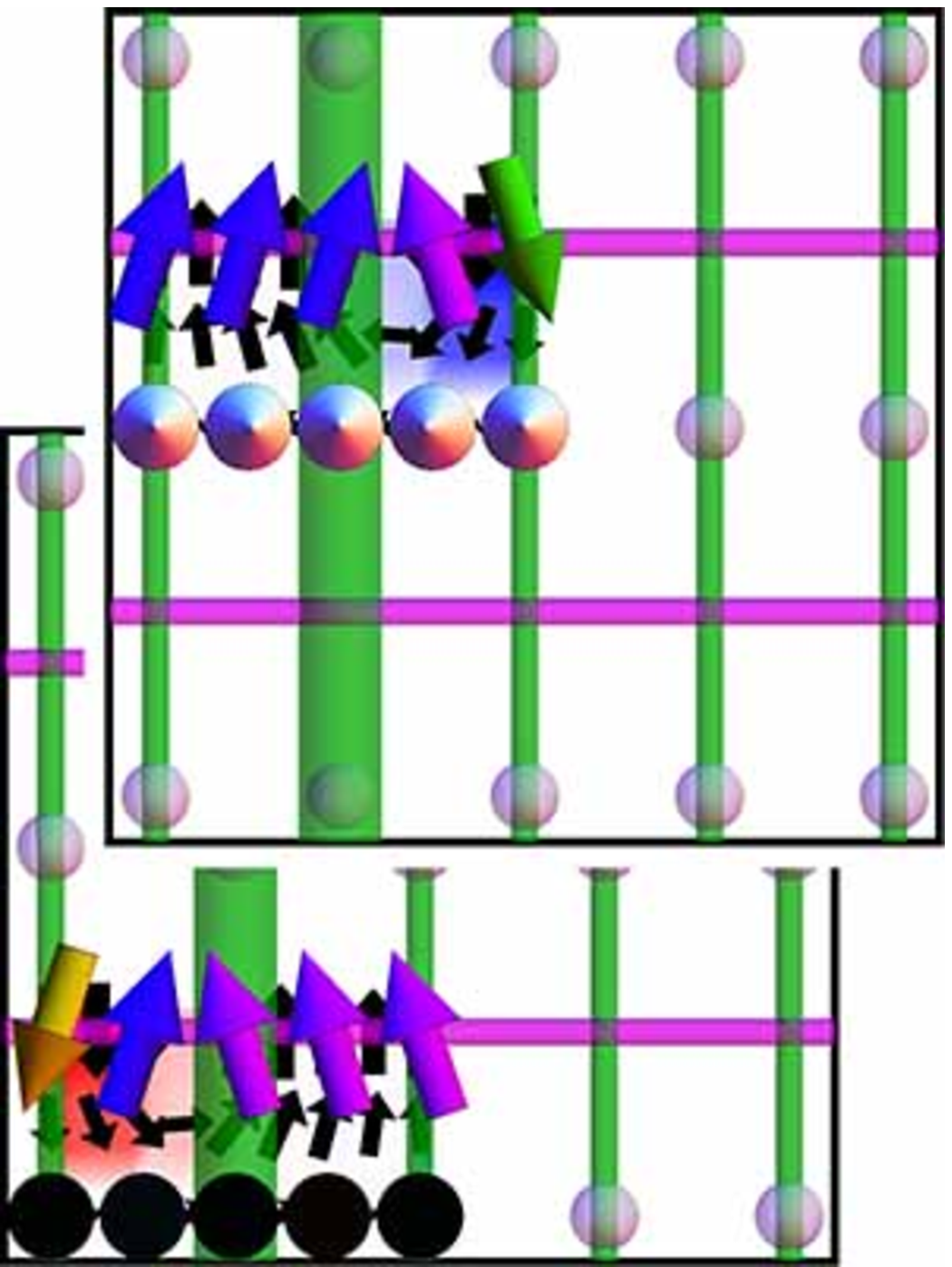}\\
\begin{sideways}
\textbf{(c)} Rotation 
\end{sideways}&
\includegraphics[width=1.5in]{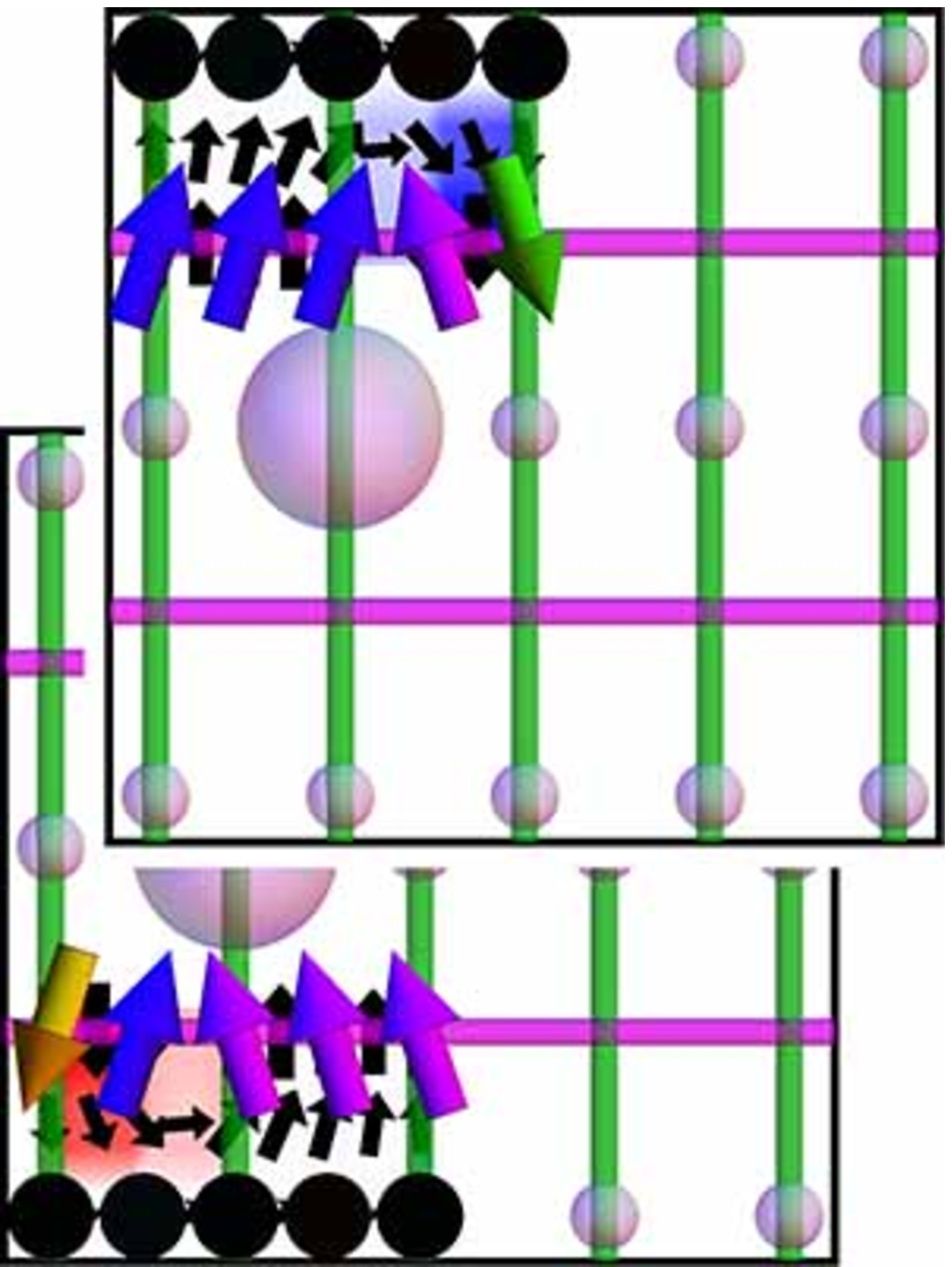}&
\begin{sideways}
\textbf{(d)} Translation 
\end{sideways}&
\includegraphics[width=1.5in]{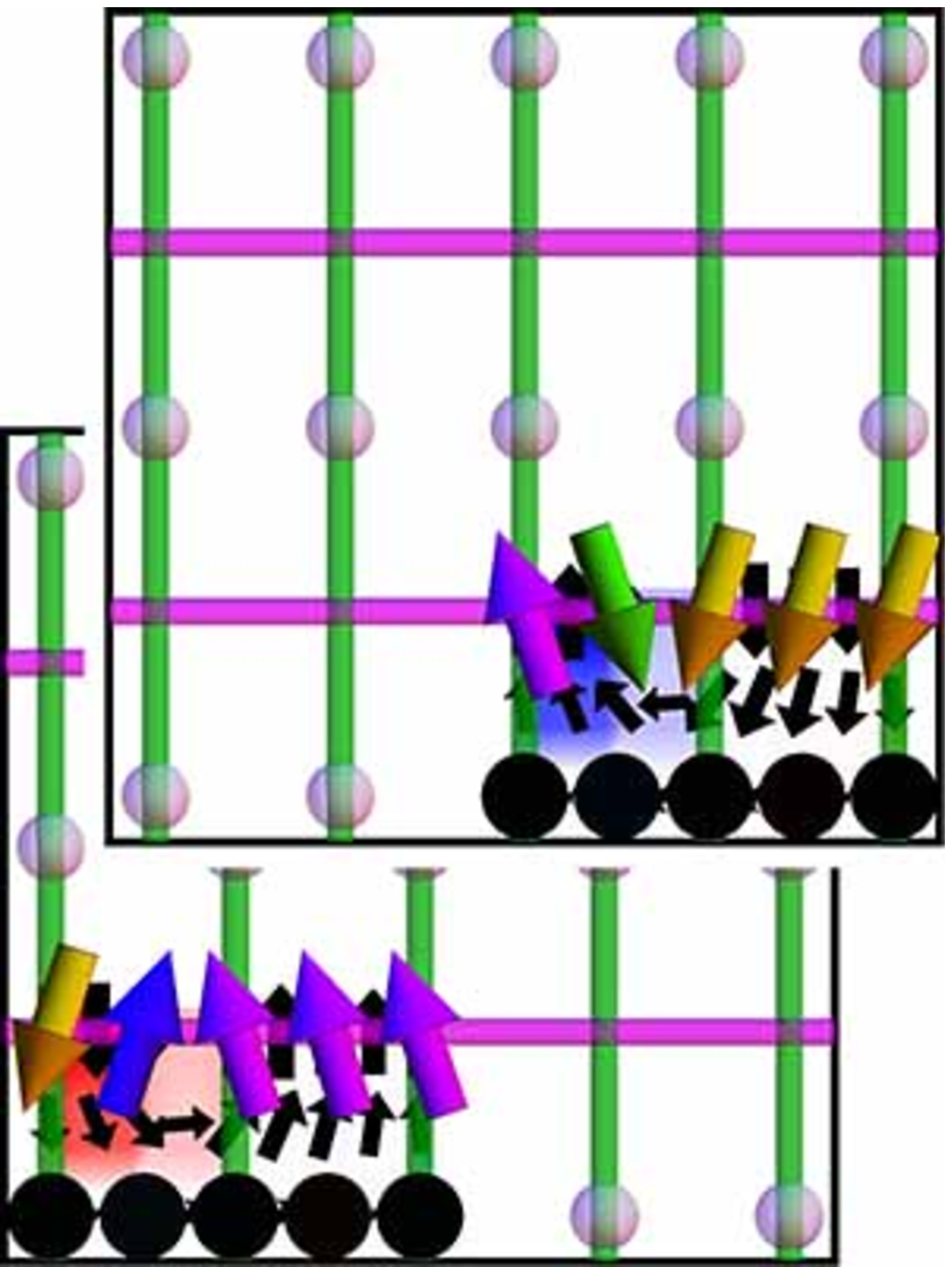}\\
\end{tabular}
\end{center}
\caption[Schematic of spin group operations]{
Figures \textbf{a}-\textbf{d} illustrates spin group
operations that combine non-trivial spin-space and real-space actions.
One element of the spin group
acts as a real-space reflection about the thick purple
mirror line combined with spin-space reflection
$\hat{n}_\parallel\rightarrow-\hat{n}_\parallel$
of the component along $\hat{B}$.
This action of this operation on the back panel is shown
in the front panel of \textbf{(a)}.
Notice spins pointing along $-\hat{B}$ below the thick 
purple line map to those along $+\hat{B}$ above.
Those on the thick purple are mapped to themselves and are perpendicular
to $\hat{B}$.
\textbf{(b)} shows a vertical translation followed by reflection 
along a glide mirror line (thick green line) combined with 
$\hat{n}_{\perp,1}\rightarrow-\hat{n}_{\perp,1}$, 
$\hat{n}_{\parallel}\rightarrow-\hat{n}_{\parallel}$
\textbf{(c)} $\pi$-rotation about a rotationd center (large white sphere)
combined with 
$\hat{n}_{\perp,1}\rightarrow-\hat{n}_{\perp,1}$,
\textbf{(d)} horizontal translation combined with 
 $\hat{n}_{\perp,1}\rightarrow-\hat{n}_{\perp,2}$
where $\hat{n}_{\perp,1}$, $\hat{n}_{\perp,2}$ are
the two components perpendicular to $\hat{B}$.
}
\label{chap:symmetry:fig:symmetry}
\end{figure}

It is also helpful to visualize the action of the group elements on
spin textures and their corresponding skyrmion densities.
For example, in Fig. \ref{chap:symmetry:fig:symmetry}a, we show
the action of a real-space reflection about the thick purple
mirror line combined with spin-space reflection
$\hat{n}_\parallel\rightarrow-\hat{n}_\parallel$
of the component along $\hat{B}$. 
The spin texture in the back panel which is entirely below
the purple line is mapped to be above the purple line in the front panel.
In addition, the spins that point along $-\hat{B}$ below the purple
line to point along $+\hat{B}$ above the purple line.  Since spins
on the purple line are mapped to themselves, consistency with the
action Eq. \ref{chap:symmetry:eq:action_x} implies the $\hat{B}$ 
component in spin-space must vanish.  This ensure continutiy of the
spin texture across the purple line.  For pure reflections about one
axis, the corresponding determinant of the real space reflection
$\det[O]$ is negative.
In addition, the determinant of the matrix describing the
spin space reflection $\hat{n}_\parallel\rightarrow-\hat{n}_\parallel$
is also negative.
Since the skyrmion density transforms with the product of the determinants,
it has the same sign going from below the thick purple
line in the back panel to above it in the front panel.

In addition to reflection about mirror lines, we also show translations
followed by reflections about glide mirror lines combined with
full spin-space inversion in Fig. 
\ref{chap:symmetry:fig:symmetry}b.  The corresponding spin group operations shows
a non-trivial combination of all three real-space translation,
real-space reflection, and spin-space inversion.  Notice it
leaves no point in real-space fixed
Fig. \ref{chap:symmetry:fig:symmetry}c shows a real-space rotation combined
with inversion of the component perpendicular to $\hat{B}$ in spin-space.
It leaves the rotation point fixed with the spin along $+\hat{B}$.
Finally, we show a translation combined with spin-space reflection
in Fig. \ref{chap:symmetry:fig:symmetry}d. 

We now begin the analysis of the constraints in Section
\ref{chap:symmetry:sec:constraints}. Recall
that a spin group is given by a choice of space group $SG$ with elements $(M,t)$, 
three-dimensional orthogonal representation $\phi$, and a
choice of a group of three-dimensional orthogonal matrices $N$ that satisfy
$\phi(M,t)^{-1}N\phi(M,t)=N$.

We first use the constraints to identify the space group $SG$.
To do this, we need to specify the Bravais lattice and point group.
The constraint (d) states that the observed spin textures directly
identify the Bravais lattice as rectangular.
From constraint (c), real-space rotation/reflection symmetry is
explicitly broken to the dihedral group $D_2$ that leaves the
magnetic field $\hat{B}$ fixed.  In general, we do not expect the spin texture
to have a higher symmetry than the Hamiltonian itself which suggests
the point group symmetry should not be larger than $D_2$.  In principal,
the point group symmetry could be spontaneously broken to a smaller
point group.  However, we assume this does not occur and take the point group
to be $D_2$.  Referring to Table \ref{chap:symmetry:tab:space_group},
we see there are a total of three space groups with a rectangular
Bravais lattice and $D_2$ point group: $p2mm$, $p2mg$, $p2gg$.

Now we turn to identifying the group $N$.  Recall $N$ has the physical
interpretation of describing the global spin-space symmetries that 
do not act on spatial degrees of freedom.  In particular, if there is a non-trivial
rotation in $N$, the spin texture then must lie along that axis.
If there is a non-trivial reflection, the spin texture must lie in 
the plane fixed by the reflection.
Constraint (e) states that spin textures cover spin space and are 
not confined to a single axis or plane.  This implies that $N$ must be the trivial
group and there are no global spin-space symmetries.

\begin{table*}
\begin{center}
\begin{tabular}{c|cc|ccc|c|c|c|c|lll}
$SG$&$BW SG$&$SG^{1/2}$&$k$&$PG_{k}$&$\psi^{PG_{k}}$&
$\phi(T_1)$&$\phi(T_2)$&$\phi(R)$&$\phi(S)$&
$a$&$b$&$E$\\
\hline
$p2mm$&$p(2a)2m'm'$&$p2mg$&$(\pi/a_1,0)$&$D_2$&$E_1$&
$--\bm{\hat{-}}$&$++\bm{\hat{+}}$&$-+\bm{\hat{-}}$&$++\bm{\hat{-}}$&
9&\textbf{\em 7}&
-0.49\\
$p2mm$&$p(2a)2mm$&$p2mm$&$(\pi/a_1,0)$&$D_2$&$A_0,B_1$&
$\bm{\hat{-}}--$&$\bm{\hat{+}}++$&$\bm{\hat{-}}-+$&$\bm{\hat{-}}-+$&
5&\textbf{\em 9}&
-0.52\\
\hline
$p2mg$&---&---&$(\pi/a_1,0)$&$D_2$&$E_1$&
$+\bm{\hat{+}}+$&$+\bm{\hat{+}}+$&$-\bm{\hat{+}}-$&$-\bm{\hat{-}}+$&
5&\textbf{\em 9}&
-0.88\\
$p2mg$&$p2'm'g'$&$p1m1$&$(\pi/a_1,0)$&$D_1$&$A_0,A_1$&
$++\bm{\hat{+}}$&$++\bm{\hat{+}}$&$++\bm{\hat{-}}$&$--\bm{\hat{+}}$&
7&\textbf{\em 9}&
-0.26\\
$p2mg$&$p(2b)2m'g'$&$p2gg$&$(\pi/a_1,0)$&$D_2$&$E_1$&
$+\bm{\hat{+}}+$&$-\bm{\hat{+}}+$&$-\bm{\hat{+}}-$&$+\bm{\hat{-}}+$&
4.2&\textbf{\em 4.5}&
-0.99\\
$p2mg$&$p(2b)2m'g'$&$p2gg$&$(0,\pi/a_2)$&$D_2$&$E_1$&
$\bm{\hat{+}}++$&$\bm{\hat{-}}--$&$\bm{\hat{-}}+-$&$\bm{\hat{+}}+-$&
\textbf{\em 9}&7&
-0.76\\
$p2mg$&$p(2b)2m'g'$&$p2gg$&$(\pi/a_1,\pi/a_2)$&$D_2$&$E_1$&
$+\bm{\hat{+}}+$&$-\bm{\hat{-}}-$&$-\bm{\hat{-}}+$&$+\bm{\hat{+}}-$&
\textbf{\em 7}&9&
-0.71\\
$p2mg$&$p(2b)2mg$&$p2mg$&$(\pi/a_1,0)$&$D_2$&$E_1$&
$++\bm{\hat{+}}$&$-+\bm{\hat{+}}$&$-+\bm{\hat{-}}$&$--\bm{\hat{+}}$&
9&\textbf{\em 9}&
-0.25\\
$p2mg$&$p(2b)2mg$&$p2mg$&$(0,\pi/a_2)$&$D_2$&$E_1$&
$\bm{\hat{+}}++$&$\bm{\hat{-}}--$&$\bm{\hat{-}}-+$&$\bm{\hat{+}}++$&
\textbf{\em 9}&3&
-0.76\\
\hline
$p2gg$&---&---&$(\pi/a_1,0)$&$D_2$&$E_1$&
$\bm{\hat{+}}++$&$\bm{\hat{+}}++$&$\bm{\hat{-}}+-$&$\bm{\hat{-}}-+$&
\textbf{\em 3}&4&
-0.89\\
$p2gg$&$p2'gg'$&$p1g1$&$(\pi/a_1,0)$&$D_1$&$A_0,A_1$&
$++\bm{\hat{+}}$&$++\bm{\hat{+}}$&$++\bm{\hat{-}}$&$--\bm{\hat{+}}$&
5&\textbf{\em 5}&
-0.85\\
\end{tabular}
\end{center}
\caption[Representations for space group generators of compatible spin groups]{

Table of spin space operations associated with the generators
of real space operations for compatible spin groups consistent with
all constraints.
For each space group $SG$ describing real space operations generated
by translations $T_1$, $T_2$, rotations $R$, and refelections $S$,
there are multiple ways to associate a real orthogonal representation of $SG$
that defines the combined spin-space operations
$\phi(T_1)$, $\phi(T_2)$, $\phi(R)$, $\phi(S)$.  Signs
indicate the diagonal entries of the corresponding matrix acting
in spin-space.

Each of these real orthogonal representations is built from either
a unitary representation of $SG$ or anti-unitary representation
or a black-white space group $BWSG$ with halving subgroup $SG^{1/2}$.
The corresponding unitary representations and anti-unitary co-representations
are specified by the wavevector $k$, wavevector point group $PG_k$,
and projective representation $\psi^{PG_k}$.  For more details, see
Appendices \ref{chap:symmetry:app:space_group}, 
\ref{chap:symmetry:app:example},
and \ref{chap:symmetry:app:table}.

Minimal energy crystalline spin textures for each resulting symmetry
group have lattice constants $a$, $b$ in units of 10 $\mu$m
for the translations $T_1$, $T_2$ and energy $E$ scaled by $g_dn_{3D}$ 
where $g_d$ is the dipolar interaction strength and $n_{3D}$ is the
peak three-dimensional density.
For $\phi$, the bold entry with a hat
indicates the component parallel to $\hat{B}$.
For real space lattice constants $a$, the bold italic entry indicates
the component parallel to $\hat{B}$.
}
\label{chap:symmetry:tab:allowed}
\end{table*}

Finally, we turn to the identification of the representation $\phi$.
The basic principle is to first enumerate all off the 
three-dimensional orthogonal representations for the space groups
$p2mm$, $p2mg$, $p2gg$.  
We use techniques described in \cite{cracknell-72,kim-99} in order to 
study two-dimensional
complex unitary representations and anti-unitary co-representations
to then analyze the needed three-dimensional real orthogonal representations.
Enumeration of these representations is the most mathematically involved 
part of the analysis and is discussed in detail in the following
appendices.
Appendix \ref{chap:symmetry:app:reps} contains a discussion
of of unitary representations, anti-unitary co-representations, 
and orthogonal representations as well as how to construct them.
Appendix \ref{chap:symmetry:app:pg} collects detailed
information about point groups in two dimensions necessary for
the construction of space group representations.
Appendix \ref{chap:symmetry:app:space_group} applies the results
of the above two appendices to the construction of
unitary representations and anti-unitary co-representations of
space groups.
Appendix \ref{chap:symmetry:app:example} presents a illustrative
example explicitly constructing the spin group for the minimal energy
spin texture shown in Fig. \ref{chap:symmetry:fig:minimal_x}.
Finally, Appendix \ref{chap:symmetry:app:table} discusses how the
compatible spin groups in Table \ref{chap:symmetry:tab:allowed}
are selected from the enumeration of all possible spin groups
in more detail.  We give a brief overview of this process
below.

After enumerating all of the representations and obtaining the associated
spin groups, we study the real-space and momentum-space actions
on both the magnetization $\hat{n}$ and skyrmion density $q$ in 
Eqs. \ref{chap:symmetry:eq:action_x} and \ref{chap:symmetry:eq:action_k}.

For a point $x$, consider spin group operations $(M,t,O)$ that leave 
$x$ fixed.  The magnetization vector $\hat{n}$ must then be left fixed
by all of the associated spin-space operations $O$.  From constraint
(b), there must be a non-trivial subspace left fixed by $O$ because
otherwise the magnetization vector would vanish at $x$.  In momentum
space, consider the wavevector $k=0$.  Similar considerations show
that for the spin group operations $(M,t,O)$ that leave $k=0$ fixed,
the spin-space operations $O$ must leave the net magnetization $\hat{n}(k=0)$
fixed.  Unlike in real space, constraint (f) implies there is no subspace
left fixed by $O$ in order to have vanishing net magnetization.
Constraint (a) implies that at least one of the $(M,t,O)$ that leave $k=0$ fixed,
must have $\det[0]\det[M]=-1$ in order to have vanishing net skyrmion
charge.

There are only 11 orthogonal representations and thus spin groups 
that satisfy 
all of the above constraints
arising from the real-space and momentum space actions.
The most general element of the spin group is given by 
Eq. \ref{chap:symmetry:eq:spin_group}.  Since
$n$ is always the identity element because $N$ is the trivial group,
we need $\phi(M,t)$ for a general element $(M,t)$ of the corresponding 
space group.
From Eq. \ref{eq:symmetry:eq:hom_gen}, we see that we only
need to specify the values of the representation for the generators
$R$, $S$ for rotation, reflections and $T_1$, $T_2$ for translations.
Table \ref{chap:symmetry:tab:allowed} gives these values for all of
the compatible spin groups. In addition, we also list the corresponding
values of the optimized lattice constants and energies obtained in the numerical
analysis of the next section.

\section{Minimal energy spin textures}

Identifying the compatible spin groups allows us to divide the space
of possible spin textures into symmetry classes.  In this section,
we describe the numerical optimization used to obtain minimal energy
spin textures.
We consider the spin texture
\begin{align}
\hat{n}(u_1,u_2)=\hat{n}(u_1t_1/N_1+u_2t_2/N_2)
\end{align}
where we take the spin texture to be in the symmetry class described by a
spin group with basis vectors $t_1$ and $t_2$.
Next we impose the spin group symmetry operations 
given by \ref{chap:symmetry:eq:action_x}.
By using the lattice of real-space translations,
we restrict our attention to the unit cell with $0\le u_i<N_i$.
This corresponds to $N_1\times N_2$ discretized points for the spin texture.

However, the number of independent points within each unit cell
is smaller due to the presence of point group operations.  
For each point $x=u_1t_1/N_1+u_2t_2/N_2$,
consider the space group elements $(M,t)$ that fix $x$.
The associated $\phi(M,t)$ in the spin group must leave $\hat{n}(x)$
invariant and gives the space of allowed $\hat{n}$ at the point $x$.
In addition, for $(M,t)$ that takes $x$ to a different point $x'$,
the magnetization at the latter point is given solely in terms of the
magnetization at the former through
$\hat{n}(x)=\hat{n}'(x')=\phi(M,t)\hat{n}(Mx+t)$ in a notation
with suppressed indices.
The independent points are given by $0\le u_i\le N_i/N'_i$ with $N_i/N'_i$ 
an integer.  For the compatible spin groups in Table 
\ref{chap:symmetry:tab:allowed} we have $N_i/N'_i=2$.

This smaller region of $N'_1\times N'_2$ points 
contained within the unit cell of $N_1\times N_2$ points
is called the fundamental region.
By specifying the spin texture within the fundamental region,
we can construct the entire spin texture via the spin group operations.
The action of the point group operations along with their
associated spin-space actions determine the spin texture within the unit
cell given its values in the fundamental region.
In particular, each element of the point group maps the fundamental
region into a distinct region within the unit cell.
This gives the ratio of the number of points in the 
fundamental region
to the number of points in the unit cell as the order or number of
group elements for the point group.  For the compatible spin groups,
the point group is $D_2$ which is of order 4.

The action of the translation operations along with their
associated spin-space actions determine the spin texture for different
unit cells.
This is shown in Fig. \ref{chap:symmetry:fig:symmetry} where the spin texture 
for coordinates 
in the lower left corner specifies the entire spin texture for all coordinates
through the symmetry group operations.

Finally, we turn to energy minimization of the resulting symmetry adapted 
discretization. 
The non-local skyrmion and dipolar interactions provides
the main difficulty which we handle via Ewald summation \cite{allen-89}.
We separate each of these interactions into short-ranged and long-ranged
contributions calculate their contributions in real and momentum space, 
respectively.  In order to approximate smooth spin textures, it is useful to 
perform an interpolation step on the discretized values before calculating
the energy.  Since the magnetization $\hat{n}$ is a unit vector living
on the sphere, this becomes a problem of spherical interpolation which
we address in detail in Appendix \ref{chap:symmetry:app:spherical}.

For each compatible spin group, we use an $8\times 8$ discretization,
fix the lattice constants $a$, $b$ and
minimize the energy $E$ with respect to the discretized spin texture
$\hat{n}(x)$.
We then minimize with respect to the lattice constants $a$, $b$.  
The results are shown in 
Table \ref{chap:symmetry:tab:allowed} for $a$, $b$ in units of 10 $\mu$m and 
$E$ scaled by the dipolar interaction energy $g_dn_{3D}$.  We check
for convergence by repeating the above procedure for a $16\times 16$ 
discretization for $a_1$, $a_2$ near the previously optimized values.

\begin{figure}
\begin{center}
\includegraphics[width=3in]{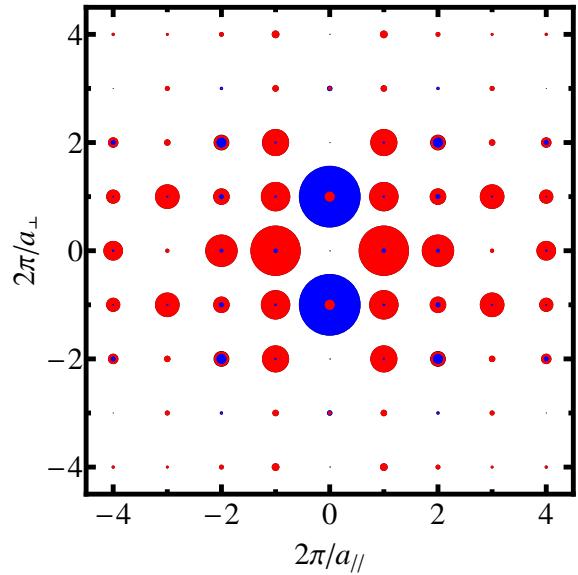}
\end{center}
\caption[Momentum space structure factor for minimal energy 
crystalline spin texture]{
Momentum space structure factor $\hat{n}(k)$ for the 
minimal energy crystalline spin texture of Fig. \ref{chap:symmetry:fig:minimal_x}.
The magnetic field $\hat{B}$ is along the horizontal axis while
lattice constants are $a_\parallel=90$ $\mu$m, $a_\perp=42$ $\mu$m.
The area of blue (red) disks is proportional to the
magnitude of components parallel (perpendicular) to 
$\hat{B}$.
}
\label{chap:symmetry:fig:minimal_k}
\end{figure}

A more refined optimization of $a$, $b$ gives the additional significant
figures for the minimal energy crystalline spin texture.  The unit cell
for this spin texture showing the magnetization $\hat{n}$, skyrmion
density $q$, and superfluid velocity $\mathbf{v}$ 
is shown in Fig. \ref{chap:symmetry:fig:minimal_x}.  
From Table \ref{chap:symmetry:tab:allowed},
notice $t_1$ ($t_2$) giving translations  perpendicular (parallel) to $\hat{B}$ 
have trivial (non-trivial) spin space operations.  This is similar to the
distinction between the unit cell and magnetic unit cell for magnetically
ordered crystals.  Fig. \ref{chap:symmetry:fig:minimal_x} plots the analog
of the magnetic unit cell.  Pure real-space translations without any
spin-space operations are sufficient to generate the rest of the spin texture.
In contrast, the unit cell corresponds to only the left (equivalently right)
half of the magnetic unit cell.  These halves are related by a spin group
operation combining a real-space translation and non-trivial spin-space operation.

This means that the magnetic unit cell lattice constants are related to the 
spin grip unit cell lattice constants by $a_\parallel=2b$ ($a_\perp=a$).  
We also plot the momentum space structure factors for components of the 
magnetization perpendicular and parallel to $\hat{B}$
in Fig. \ref{chap:symmetry:fig:minimal_k}.

\section{Discussion}
With the symmetry analysis and energy minimization completed, 
we now discuss the structure of the resulting crystalline
spin textures.  We first focus on the minimal energy spin texture
shown in Figs. \ref{chap:symmetry:fig:minimal_x}, 
\ref{chap:symmetry:fig:minimal_k}.  From
the momentum space spin structure factors in Fig. \ref{chap:symmetry:fig:minimal_k},
spin components parallel (perpendicular) to $\hat{B}$
have weight concentrated at wavevectors perpendicular (parallel) to $\hat{B}$.
This anisotropy in the structure factor weights maximizes the gain in the
dipolar interaction energy in Eq. \ref{chap:symmetry:eq:dipolar_effective}.
This pattern is also consistent with the characteristic cross-like structure 
for observed spin structure factors
\cite{dsk-08,dsk-09}.
It also agrees with the pattern of unstable modes obtained from
a dynamical instability analysis 
of the uniform state \cite{cherng-09,ueda-09}.

Notice that such a spin texture has a non-vanishing skyrmion density $q$ as 
shown in Fig. \ref{chap:symmetry:fig:minimal_x}.  
This follows from Eq. \ref{chap:symmetry:eq:q} showing $q\ne0$ 
when orthogonal components of $\hat{n}$ vary along orthogonal directions.
Since the vorticity of the superfluid velocity $\mathbf{v}$ is given by $q$, 
this implies the presence of persistent, circulating superfluid currents.

Consider the components $\hat{n}_\parallel$ ($\hat{n}_{\perp}$) parallel 
(perpendicular) to $\hat{B}$ separately in the region 
$0\le x_\parallel\le a_\parallel/2$, $0\le x_\perp\le a_\perp/2$ of
Fig. \ref{chap:symmetry:fig:minimal_x} where $x_\parallel$ ($x_\perp$) are coordinates
parallel (perpendicular) to $\hat{B}$.
Symmetry operations give the spin texture in all other regions.
We can characterize the behavior of parallel components as
$\hat{n}_\parallel\sim\cos(k_\perp x_\perp)$ 
varying over the entire range $\pm1$ while perpendicular components
$\hat{n}_{\perp,1}+i\hat{n}_{\perp,2}\sim
\sqrt{1-\hat{n}_\parallel^2}\exp(ik_\perp x_\perp)$ 
have a spiral winding in regions between $\hat{n}_\parallel=\pm1$.  
The dipolar interaction favors this configuration
and gives rise to a non-vanishing 
skyrmion density $q$ and superfluid velocity $\mathbf{v}$.

In the companion paper, we showed spin textures of this form arise naturally
even in the absence of dipolar interactions as non-trivial analytical solutions of
the effective theory with spin stiffness and skymrion interactions.  There
they have an interpretation as neutral stripe configurations of skyrmions
and anti-skyrmions.  Turning on dipolar interactions makes such
solutions more stable compared to the uniform state.

In conclusion, we have considered the low-energy effective theory for
dipolar spinor condensates.  The resulting non-linear sigma model
describes the dynamics of the magnetization and includes spin stiffness,
skyrmion interaction, and dipolar interaction terms. 
A systematic analysis of symmetry
operations containing combined real space and spin space actions allows us
to classify the allowed symmetry groups consistent with 
non-trivial theoretical and experimental constraints on possible spin textures.
Possible ground states describing neutral collections of topological 
skyrmions carrying persistent superfluid currents are obtained by 
minimizing the energy within each symmetry class.

\begin{acknowledgements}
We thank 
D. Stamper-Kurn, M. Vengalattore, G. Shlyapnikov, S. Girvin, T.-L. Ho,
A. Lamacraft,  and M. Ueda for stimulating discussions. 
This work was supported by a NSF Graduate Research
Fellowship, 
NSF grant DMR-07-05472,  AFOSR Quantum Simulation MURI, AFOSR MURI on
Ultracold Molecules, DARPA OLE program, and Harvard-MIT CUA.
\end{acknowledgements}

\appendix

\section{\label{chap:symmetry:app:reps}Group theory and representation theory}

To provide background for the analysis of spin groups, we review here
aspects of group theory and representation theory.
Beginning with general definitions
for group and representations, we then discuss
unitary representations of groups and
their generalization to projective unitary
representations.  Then we consider how projective unitary representations are used
in the construction of the unitary representations of a group
with a normal abelian subgroup.
Next we present anti-unitary co-representations and how to construct them
from the unitary representations of a halving subgroup.  Finally, we analyze
the real orthogonal representations relevant for spin groups and how to obtain
them from unitary representations and anti-unitary co-representations.
For more details on group theory, see Ref. \cite{artin-91}. 
For applications of group theory to the study of space groups,
see Refs. \cite{cracknell-72,kim-99}.

\subsection{Group theory}
A group $G$ is a set of elements $g$ with a binary operation 
$G\times G\rightarrow G$ usually called multiplication
satisfying the axioms of closure $g_1 g_2 \in G$, 
associativity $(g_1g_2)g_3=g_1(g_2g_3)$, identity $\mathbf{1} g=g$ for 
the identity element $\mathbf{1}$, and inverse $g^{-1} g=\mathbf{1}$ for
the inverse element $g^{-1}$. $|G|$ is the order or number of elements in the group.

A subgroup $H$ of a group $G$ is a subset of elements $h$ in $G$ that also
form a group under multiplication in $G$.  A normal subgroup $N$ of a group $G$
is a subgroup that is left fixed by conjugation $g^{-1}Ng=N$ for all elements
$g$ in $G$.  An abelian group is a group that is commutative $g_1g_2=g_2g_1$.

When $H$ is a subgroup of $G$, the equivalence relation $g_1\sim g_2$ for
$g_1^{-1}g_2 \in H$ divides $G$ into distinct equivalence classes.  For
representatives $r_i$, the left cosets $r_iH$ form the equivalence classes 
$G=\cup_i r_i H$.  Similarly, the right cosets $Hg_i$ also form the equivalence
classes $G=\cup_i H r_i$.  A normal subgroup has the same left and right
cosets $r_iH=Hr_i$.  In this case, the left
(equivalently the right) 
cosets form a group called the quotient group $G/H$ with multiplication
defined by $r_1 r_2 = r_3(r_1,r_2)$ where $r_3(r_1,r_2)$ 
is defined as the coset representative
that satisfies $r_1 r_2 H = r_3(r_2,r_2) H$.

A group homomorphism $\phi$ from group $G$ to group $H$ is a map $G\rightarrow H$
that is compatible with both of the group multiplications.  This means that
$\phi$ satisfies
the homomorphism condition $\phi(g_1)\phi(g_2)=\phi(g_1 g_2)$. 
The kernel $\text{ker}(\phi)$ consists of the elements of $G$ that map to the
identity element in $H$, $\phi(g)=\mathbf{1}$.  The kernel is a normal subgroup
of $G$.  The image $\text{im}(\phi)$ consists of the elements of $H$ that
occur for some element $g$ of $G$.  The image is a subgroup of $H$.
A surjective map $\phi$ is equivalent to $\text{im}(\phi)=G$ being the entire group $H$.
An injective map $\phi$ is equivalent to $\text{ker}(\phi)=\mathbf{1}$ 
being the trivial group
consisting of the identity element alone.
A bijective map $\phi$ is called an isomorphism.

\subsection{Group representations}
A representation $\phi$ is a homomorphism from a group $G$ to 
the group of linear transformations on some finite dimensional vector space $V$.
This means that for each group element $g$ in $G$, $\phi(g)=M$
where $M$ is an invertible matrix.  Here $\phi$ is subject to the
homomorphism condition
\begin{align}
\phi(g_1)\phi(g_2)=\phi(g_1g_2)
\end{align}
Two representations $\phi$, $\phi'$ are equivalent if there is a fixed matrix $S$
such that
\begin{align}
S^{-1}\phi(g)S=\phi'(g)
\end{align}
for all group elements $g$ in $G$.
A representation is irreducible if the action of the group through $\phi(g)$
leaves no non-tirival subspace fixed.
The primary goal of group representation theory is to classify 
and construct all of the inequivalent
irreducible representations.

\subsection{Unitary representations}
A unitary represenation $\phi_U$ is a homomorphism from a group $G$ to
the group of finite-dimensional unitary transformations.  
This means that for each
group element $g$ in $G$, $\phi_U(g)=U(G)$ is a finite-dimensional 
complex unitary matrix satisfying
\begin{align}
U(g)^{-1}=U(g)^{\dagger}
\end{align}
where $\dagger$ denotes the adjoint or complex conjugate transpose.
Here, $\phi_U$ is subject to the homomorphism condition
\begin{align}
\phi_U(g_1)\phi_U(g_2)=\phi_U(g_1g_2)
\end{align}
for each element $g_1$, $g_2$ in the group $G$.
In particular, this implies 
\begin{align}
U(g_1)U(g_2)=U(g_1g_2)
\end{align}
for the unitary matrices associated to 
the elements $g_1$, $g_2$ in the group $G$.

\subsection{Projective unitary representations}
A projective unitary representation $\psi_U$ is a homomorphism from a group $G$
to the group of finite-dimensional projective unitary 
transformations.  Projective unitary transformations
that only differ by multiplication by a complex scalar are considered the same.  
In constrast, unitary
transformations the differ by multiplication by a complex scalar are distinct.
This means that for each group element $g$ in $G$, $\psi_U(g)=U(G)$ is
a finite-dimensional complex unitary matrix satisfying
\begin{align}
U(g)^{-1}=U(g)^{\dagger}
\end{align}
where $\dagger$ denotes the adjoint or complex conjugate transpose.
Here, $\psi_U$ is subject to the projective homomorphism condition
\begin{align}
\label{chap:symmetry:eq:proj}
\psi_U(g_1)\psi_U(g_2)=\lambda(g_1,g_2)\psi_U(g_1g_2)
\end{align}
where $\lambda(g_1,g_2)$ is the factor system for the projective representation
and is a complex scalar for all elements $g_1$, $g_2$ in the group $G$.
In particular, this implies 
\begin{align}
U(g_1)U(g_2)=\lambda(g_1,g_2)U(g_1g_2)
\end{align}
for the unitary matrices associated to 
the elements $g_1$, $g_2$ in the group $G$.

The factor system $\lambda(g_1,g_2)$ 
is subject to the associativity
condition
\begin{align}
\lambda(g_1,g_2)\lambda(g_1g_2,g_3)&=\lambda(g_1,g_2g_3)\lambda(g_1,g_2)
\end{align}
Two projective representations $\psi'_U,\lambda'$ and $\psi_U,\lambda$ are 
projectively
equivalent if there exists a fixed matrix $S$ and non-zero complex
scalar function $l(g)$ such that
\begin{align}
S^{-1}\psi_U(g)S/l(g)=\psi'_U(g)
\label{chap:symmetry:eq:proj_equiv}
\end{align}
from which we can see that the factor systems are related by
\begin{align}
\lambda'(g_1,g_2)=\frac{\lambda(g_1,g_2)}{l(g_1)l(g_2)}
\end{align}
Projective equivalence divides the
projective representations of a group into equivalence classes.
From each equivalence class, we can choose a normalized and standard factor system
representative subject to the normalization and standardization conditions
\begin{align}
|\lambda(g_1,g_2)|&=1,&
\lambda(g,\mathbf{1})&=\lambda(\mathbf{1},g)=\lambda(\mathbf{1},\mathbf{1})=1
\end{align}
where $\mathbf{1}$ is the identity element.

\subsection{Induced and subduced representations}
Consider a group $G$ with a subgroup $H$ of index $I=|G|/|H|$ where recall
$|G|$ denote the order or number of elements in group $G$.
The left coset decomposition of $G$ by $H$ is given by
\begin{align}
G=\cup_i r_i H
\end{align}
where $r_1\ldots r_I$ are left coset representatives.
For a $N\times N$ dimensional unitary representation $\phi^H$ of the subgroup $G$, 
the induced representation $\phi^{H\uparrow G}$ of the group $G$ is a 
$IN\times IN$ dimensional unitary representation 
\begin{align}
\phi^{H\uparrow G}(g)_{ij}=\sum_{h\in H} \phi^H(h)\delta(h,r_i^{-1}g r_j)
\end{align}
where $\delta$ is the Kronecker delta function.  
In the notation above,
when $r_i^{-1} g r_j=h$, the $i$ row and $j$ column with $1\le i,j\le I$ of $\phi^G(g)$ 
consists of the $N\times N$ matrix $\phi^H(h)$.

Given a unitary
representation of
$\phi^G$ of $G$, the subduced representation $\phi^{G\downarrow H}$
is given by 
\begin{align}
\phi^{G\downarrow H}(h)=\phi^{G}(h)
\end{align}
and is a unitary representation of $H$ which corresponds to restriction 
to the elements $h$ of $H$ for $\phi^G$.
The induced representation $\phi^G$ gives a unitary
representation of $G$ from
a unitary representation $\phi^H$ of a subgroup $H$.  

\subsection{\label{chap:symmetry:subsec:little}Little groups and small representations}

Given an irreducible unitary representation $\phi^H$ of a subgroup $H$ of $G$, 
the little group $G_{\phi^H}$ is the largest subgroup of $G$ that 
leaves $\phi^H$ fixed under conjugation.  This means that $G_{\phi^H}$ consists
of all elements $g$ in $G$ for which $\phi^H(g^{-1}hg)=\phi^H(h)$ is true
for all elements of $h$ in $H$.  From this, it is clear 
that $H$ is a subgroup of $G_{\phi^H}$.

A small representation 
$\phi^{G_{\phi^H}}$ of the little group $G_{\phi^H}$ is a unitary representation
of $G_{\phi^H}$ that subduces to $\phi^H$
\begin{align}
\phi^{G_{\phi^H}\downarrow H}(h)=\phi^H(h)
\end{align}

Assume that $\phi^H$ is an irreducible unitary representation of $H$.
For a small representation $\phi^{G_{\phi^H}}$ of the little group $G_{\phi^H}$,
consider the induced representation $\phi^{G_{\phi^H}\uparrow G}$ of the group $G$.
This representation is an irreducible unitary representation of $G$.
Morever, all the irreducible unitary representations of $G$ arise in this way.

Thus we see that the small representations of  the little group are a crucial
step in the construction of inequivalent irreducible unitary representations
of a group $G$ from the inequivalent irreducible unitary representations
of a subgroup $H$.  This construction is feasible only when the small
representations of the little group can be obtained.  One case where this
is the case is when $H$ is both a normal and abelian subgroup of $G$.

Suppose that $H$ is both a normal and abelian subgroup of the group $G$,
$\phi^H$ is an irreducible unitary representation of $H$, 
$G_{\phi^H}$ is the little group, and 
$\phi^{G_{\phi^H}}$ is a small representation of $G_{\phi^H}$.

Using the definition of the little group  $G_{\phi^H}$
and $H$ a normal subgroup of $G$, we see that $H$ is also a normal
subgroup of $G_{\phi^H}$.  In particular, the quotient group
$G_{\phi^H}/H$ is a subgroup of the quotient group $G/H$.
For the left coset decomposition 
$G_{\phi^H}=\cup_i r_i H$, the quotient group $G_{\phi^H}/H$ 
has the multiplication law $r_1 r_2=r_3(r_1,r_2)$ where $r_3(r_1,r_2)$ is
the coset representative satisfying $r_1r_2H=r_3(r_1,r_2)H$.  Note that
while $r_1 r_2 = r_3(r_1,r_2)$ holds in the quotient group $G_{\phi^H}/H$,
only the weaker relation $r_1 r_2 = r_3(r_1,r_2) h_3(r_1,r_2)$ 
for some element $h_3(r_1,r_2)$ in $H$
holds in the group $G_{\phi^H}$ itself.

Next consider the homomorphism relation 
$\phi^{G_{\phi^H}}(r_1)\phi^{G_{\phi^H}}(r_2)=\phi^{G_{\phi^H}}(r_1r_2)$.
By using $r_1 r_2 = r_3(r_1,r_2) h_3(r_1,r_2)$ and the homomorphism relation
$\phi^{G_{\phi^H}}(r_3(r_1,r_2) h_3(r_1,r_2))=
\phi^{G_{\phi^H}}(r_3(r_1,r_2))\phi^{G_{\phi^H}}(h_3(r_1,r_2))$ we find
\begin{align}
\label{chap:symmetry:proj1}
\phi^{G_{\phi^H}}(r_1)\phi^{G_{\phi^H}}(r_2)=
\phi^{G_{\phi^H}}(r_3(r_1,r_2))\phi^{G_{\phi^H}}(h_3(r_1,r_2))
\end{align}
Since $\phi^{G_{\phi^H}}$ is a small representation,
$\phi^{G_{\phi^H}}(h_3(r_1,r_2))=\phi^H(h_3(r_1,r_2))$ is a complex scalar.

Finally, we recognize
\begin{align}
\label{chap:symmetry:proj2}
\phi^{G_{\phi^H}}(r_1)\phi^{G_{\phi^H}}(r_2)=
\phi^H(h_3(r_1,r_2))\phi^{G_{\phi^H}}(r_3(r_1,r_2))
\end{align}
as Eq. \ref{chap:symmetry:eq:proj} for the
defining relation for a projective unitary representation.
Here the multiplication law is $r_1 r_2 = r_3(r_1,r_2)$ in the quotient group
$G_{\phi^H}/H$ and the factor system is defined by 
$\phi^H(h_3(r_1,r_2))$ where $r_1 r_2 = r_3(r_1,r_2) h_3(r_1,r_2)$ in the group 
$G_{\phi^H}$.

To summarize, when $H$ is an abelian and normal subgroup of $G$, the
inequivalent irreducible unitary representations can be constructed
as follows.  Find the inequivalent irreducible unitary representations
$\phi^H$ of $H$.  Divide the irreducible unitary representations 
$\phi^H$ into equivalence classes according to the relation 
$\phi^H_1\sim\phi^H_2$ if $\phi^H_1(h)=\phi^H_2(g^{-1}Hg)$ where $g$ is 
an element of $G$.

For each equivalence class find the little group $G_{\phi^H}$ consisting
of elements $g$ that leave $\phi^H(h)=\phi^H(g^{-1}hg)$ fixed.
Consider the quotient group $G_{\phi^H}/H$ which 
is itself a subgroup of the quotient group $G/H$.
Here the multiplication law is $r_1 r_2 = r_3(r_1,r_2)$ in  
$G_{\phi^H}/H$ and $r_1 r_2 = r_3(r_1,r_2) h_3(r_1,r_2)$ in $G_{\phi^H}$
for $h_3(r_1,r_2)$ in $H$.
Find the irreducible projective unitary representations
$\psi^{G_{\phi^H}/H}$ that are
projectively equivalent to the factor system $\phi^H(h_3(r_1,r_2))$.

Each of these  projective unitary representations 
$\psi^{G_{\phi^H}/H}$ of $G_{\phi^H}/H$ can be extended to 
a unitary representation of $G_{\phi^H}$.
This can be seen as follows.  For an arbitrary element $g$ of
$G_{\phi^H}$, we can write the left coset decomposition 
$g=rh$ for some coset representative $r$ and $h$ an element
of $H$.  By taking $\psi^{G_{\phi^H}/H}(g)=\psi^{G_{\phi^H}/H}(r)\phi^H(h)$,
we can use Eqs. \ref{chap:symmetry:proj1} and \ref{chap:symmetry:proj2}
to show that $\psi^{G_{\phi^H}/H}(g_1g_2)=\psi^{G_{\phi^H}/H}(g_1g_2)$ 
satisfies the homomorphism relation.

Each of the induced
representations $\psi^{G_{\phi^H}/H\uparrow G}$ is then an
irreducible unitary representation of $G$.  
Each of these
irreducible unitary representations is inequivalent for $\phi^H$
taken from different equivalence classes under the 
$\phi^H_1\sim\phi^H_2$ if $\phi_1^H(h)=\phi_2^H(g^{-1}Hg)$ for some $g$ 
an element of $G$.  By using all of the equivalence classes,
all of the inequivalent irreducible unitary
representations of $G$ are obtained.

\subsection{\label{chap:symmetry:app:au}Anti-unitary co-representations}

In order to have anti-unitary co-represenations, the group $G$ must have
a halving subgroup $H$.  This means that $H$ is an index two
subgroup $|G|/|H|=2$ of the group $G$.  In particular, $H$ is a normal
subgroup and the quotient group $G/H$ consists of two elements:
the identity element $\mathbf{1}$ of $G$ and a left coset representative
$z$.  The group $G$ can be written as $G=H \cup zH$ with
$z$ satisfying
\begin{align}
\label{chap:symmetry:eq:z}
z&\notin H,& z^2&\in H,& z^{-1}hz&\in H
\end{align}
An anti-unitary co-representation $\phi_{AU}$ is a homomorphism from a group
$G$ to the group of complex linear and anti-linear unitary transformations.
This means that for each group element $h$ in $H$, $\phi_{AU}(h)=U(h)$ 
and for the element $z$ in $G$, $\phi_{AU}(z)=U(z)\Theta$ where
$U(h)$ and $U(z)$ are unitary matrices satisfying
\begin{align}
U(h)^{-1}&=U(h)^{\dagger}, U(z)^{-1}=U(z)^{\dagger}
\end{align}
and  $\Theta$ is the complex conjugation operator.
Here $\phi_{AU}$ is subject to the homomorphism condition
\begin{align}
\phi_{AU}(g_1)\phi_{AU}(g_2)=\phi_{AU}(g_1g_2)
\end{align}
In particular this implies
\begin{align}
\nonumber
U(h_1)U(h_2)&=U(h_1h_2)&
U(z)U(z)^*&=U(zz)\\
U(h)U(z)\Theta&=U(hz),&
U(z)U(h)^*\Theta&=U(zh)
\end{align}
for the unitary matrices associated to the elements
$h$, $h_1$, $h_2$ of the group $H$ and element $z$ of the group $G$.
For a more detailed discussion of unitary representations
and anti-unitary co-representations, see Ref. \cite{herbut-80}.

We now discuss how to construct the inequivalent irreducible 
anti-unitary co-representations $\phi^G_{AU}$ of $G$ from the
inequivalent irreducible unitary representations $\phi^H_U$ of
the halving subgroup $H$.  From Eq. \ref{chap:symmetry:eq:z}, we know
that $z^{-1}Hz=H$.  This implies that since $\phi^H_U(h)$ is an irreducible
unitary representation with $h$ an element of the group $H$,
$\phi^H_U(z^{-1}hz)^*$ with $*$ the complex conjugate is also an irreducible
unitary representation of $H$.  In particular, this means that there is
a unitary matrix $Z$ such that 
\begin{align}
\phi^H_U(z^{-1}hz)^* = Z^\dagger\varphi^H_U(h)Z
\end{align}
for some irreducible unitary representation $\varphi^H_U$ of $H$.

From Eq. \ref{chap:symmetry:eq:z}, notice $z^2$ is also an element of $H$.
There are three cases two consider.  The first case (1) is when
$\phi^H_U$ and $\varphi^H_U$ are inequivalent.  The irreducible
anti-unitary co-representation of $G$ is given by
\begin{align}
\nonumber
\phi^G_{AU}(h)&=U(h)=\begin{bmatrix}
\phi^H_U(h)&0\\
0&\varphi^H_U
\end{bmatrix},\\
\phi^G_{AU}(h)&=U(z)\Theta=\begin{bmatrix}
0&\phi^H_U(z^2)Z^T\\
Z&0
\end{bmatrix}\Theta
\end{align}
where $T$ denotes the transpose and $\Theta$ is the complex
conjugation operator.

The second case (2a) is when $\phi^H_U$ and $\varphi^H_U$ are equivalent
and $ZZ^*=-\phi^H_U(z^2)$.  The irreducible
anti-unitary co-representation of $G$ is given by
\begin{align}
\nonumber
\phi^G_{AU}(h)&=U(h)=\begin{bmatrix}
\phi^H_U(h)&0\\
0&\phi^H_U
\end{bmatrix},\\
\phi^G_{AU}(h)&=U(z)\Theta=\begin{bmatrix}
0&-Z\\
Z&0
\end{bmatrix}\Theta
\end{align}
where $\Theta$ is the complex conjugation operator.

The third case (2b) is when $\phi^H_U$ and $\varphi^H_U$ are equivalent
and $ZZ^*=+\phi^H_U(z^2)$.
The irreducible
anti-unitary co-representation of $G$ is given by
\begin{align}
\phi^G_{AU}(h)&=U(h)=\phi^H_U(h),&
\phi^G_{AU}(h)&=U(z)\Theta=Z\Theta
\end{align}
where $\Theta$ is the complex conjugation operator.

Notice that if $\phi_U^H$ is a $N\times N$ dimensional unitary representation,
then $\phi^G_{AU}$ is a $2N\times 2N$ dimensional anti-unitary co-representation
for cases (1) and (2a) while it is a $N\times N$ dimensional anti-unitary 
co-representation for case (2b).  
All of the inequivalent anti-unitary co-representations $\phi^G_{AU}$ of $G$
are obtained by using the above procedure once for each pair
of type (3) inequivalent irreducible unitary representations $\phi^H_{U}$,
$\varphi^H_{AU}$ of $H$ and once for each type (2a) or (2b) 
inequivalent irreducible unitary representation $\phi^H_{U}$ of $H$

The construction of the unitary matrix $Z$ is described in Ref. \cite{herbut-80}.
Consider the projectors
\begin{align}
P_i=\frac{N}{|H|}\sum_{h}\phi_U^H(h)_{1i}\phi_U^H(z^{-1}hz)^T
\end{align}
where $N\times N$ is the dimensionality of $\phi_U^H$, $|H|$ is the order
or number of elements in the group $H$, and $\phi_U^H(h)_{1i}$ is the $(1,i)$
scalar matrix element of $\phi_U^H(h)_{1i}$.  Let $x$ be the unique normalized 
column eigenvector $x$ with eigenvalue one for $P_1$.  Then the $i$ row of $Z$
is given by $x^\dagger P_i^\dagger$.

\subsection{\label{chap:symmetry:app:orth}Real orthogonal representations}

A real orthogonal representation $\phi_O$ is a homomorphism from a group $G$ 
to the group of linear orthogonal transformations.
This means that for each group element $g$ in $G$, $\phi_O(g)=O$ where $O$ is
a finite-dimensional real orthogonal matrix 
satisfying
\begin{align}
O(g)^*&=O(g),&O(g)^{-1}=O(g)^T
\end{align}
where $*$ denotes complex conjguation and $T$ denotes the transpose.
Here, $\phi_O$ is subject to the homomorphism condition
\begin{align}
\phi_O(g_1)\phi_O(g_2)=\phi_O(g_1g_2)
\end{align}
for each element $g'$, $g$ in the group $G$.
In particular, this implies 
\begin{align}
O(g_1)O(g_2)=O(g_1g_2)
\end{align}
for the orthogonal matrices associated to the
elements $g'$, $g$ in the group $G$

We are primarily interested in the 
three-dimensional real orthogonal representations of space groups
for the analysis of spin groups.
Luckily, the three-dimensional real orthogonal representations 
can be easily obtained from the two-dimensional
complex unitary representations and anti-unitary co-representations.
Physically, this corresponds to using two component complex unit spinors
to construct three component real vectors.  Mathematically, it corresponds to
the 2-to-1 homomorphism from $SU(2)$ to $SO(3)$.

When $U$ is a two-dimensional complex unitary matrix
\begin{align}
O^{ij}_{U}(U)&=\frac{1}{2}\text{Tr}
\left[\sigma^iU^\dagger\sigma^jU\right]
\end{align}
is a three-dimensional real orthogonal matrix. 
Similarly, when $U\Theta$ is a two-dimensional complex anti-unitary
matrix
\begin{align}
O^{ij}_{AU}(U)&=\frac{1}{2}\text{Tr}
\left[\sigma^iU^T(\sigma^j)^T U^*\right]
\end{align}
is a three-dimensional real orthogonal matrix.
In both of the above, $\sigma^i$ are the Pauli matrices.
Using the completeness relation for Pauli matrices
\begin{align}
\sum_i\sigma^{i}_{\alpha\beta}\sigma^{i}_{\gamma\delta}=
2\delta_{\alpha\delta}\delta_{\beta\gamma}-\delta_{\alpha\beta}\delta_{\gamma\delta}
\end{align}
we can show that
\begin{align}
\nonumber
O^{ij}_{U}(U_1)O^{ij}_{U}(U_2)&=O^{ij}_{U}(U_1U_2),&\\
\nonumber
O^{ij}_{AU}(U_1)O^{ij}_{AU}(U_2^*)&=O^{ij}_{U}(U_1U_2)\\
\nonumber
O^{ij}_{U}(U_1)O^{ij}_{AU}(U_2)&=O^{ij}_{AU}(U_1U_2),&\\
O^{ij}_{AU}(U_1)O^{ij}_{U}(U_2^*)&=O^{ij}_{AU}(U_1U_2),&
\end{align}
satisfies that appropriate homomorphism relations.

\section{\label{chap:symmetry:app:pg}Cyclic, dihedral, and double dihedral groups}
In two dimensions, point groups are either cyclic $C_n$ or dihedral $D_n$.
Here we discuss the structure of these groups, their subgroups,
their inequivalent irreducible unitary representations, and their
projectively inequivalent irreducible projective unitary representations.

Cyclic groups have generators that satisfy $r^n=s=\mathbf{1}$ with $n$ 
group elements $r^m$ where $m=0\ldots n-1$ and $\mathbf{1}$ is the
identity element.
These groups are abelian and the inequivalent irreducible
complex unitary representations are 
one-dimensional and labeled by an integer $\mu=0\ldots n-1$.
For the generator $r$, the representation is given by
\begin{align}
\phi^{C_n}_\mu(r)=\exp(2\pi i \mu/n),& 
\end{align}
with the homomorphism relation $\phi^{C_n}_\mu(r^m)=\phi^{C_n}_\mu(r)^m$
specifying the representation for the entire group.
The projective inequivalent irreducible 
projective unitary representations of $C_n$ are projectively equivalent to 
the complex unitary representations of $C_n$ with trivial factor system.

For the cyclic group $C_n$ of order $n$, 
the subgroups are also cyclic $C_p^n$ of order $p$ where $p$ is a divisor of
$n$.  There are $p$ elements of $C_p^n$ given by the elements
$r^{qn/p}$ of the group $C_n$ for $q=0\ldots p-1$.  The left coset decomposition
of the group $C_n$ is given by $C_n=\cup_r r^u C_p^m$ with left coset
representatives $r^u$ where $u=0\ldots n/p-1$.

Dihedral groups have generators that satisfy $r^n=s^2=\mathbf{1}$ with $2n$
group elements $r^m s^t$ with $m=0\ldots n-1$ and $t=0,1$ and $\mathbf{1}$ is the
identity element.
Such groups are non-abelian except for $n\le 2$.
For the generators $r$, $s$, the inequivalent irreducible unitary
representations are given by
Table \ref{chap:symmetry:tab:dn_reps}
with the homomorphism relation 
$\phi^{D_n}_{Rep}(r^ms^t)=\phi^{D_n}_{Rep}(r)^m\phi^{D_n}_{Rep}(s)^t$
specifying the representation for the entire group.

\begin{table}
\begin{center}
\begin{tabular}{l|l|ll}
&Rep.&$r$&$s$\\
\hline
$n$ odd&$A_0$&$+1$&$+1$\\
&$A_1$&$+1$&$-1$\\
&$E_\mu$&$\exp(2\pi i \mu \sigma^y/n)$&$\sigma^z$\\
\hline
$n$ even&$A_0$&$+1$&$+1$\\
&$A_1$&$+1$&$-1$\\
&$B_0$&$-1$&$+1$\\
&$B_1$&$-1$&$-1$\\
&$E_\mu$&$\exp(2\pi i \mu \sigma^y/n)$&$\sigma^z$
\end{tabular}
\end{center}
\caption[Inequivalent irreducible unitary representations 
of the dihedral group $D_n$]{The 
inequivalent irreducible unitary representations $\phi^{D_n}_{Rep}$ 
for the generators $r$, $s$ of the dihedral group $D_n$.
Here $\mu=1\ldots n/2-1$ for $n$ even, $\mu=1\ldots (n-1)/2$ for $n$ odd, and
$\sigma^i$ are the Pauli matrices.}
\label{chap:symmetry:tab:dn_reps}
\end{table}

For the dihedral group $D_n$ of order $n$, the subgroups are one of two
types.  The first is cyclic 
$C_p^n$ of order $p$ where $p$ is a divisor of
$n$.  There are $p$ elements of $C_p^n$ given by the elements
$r^{qn/p}$ of the group $D_n$ for $q=0\ldots p-1$.  The left coset decomposition
of the group $D_n$ is given by $D_n=\cup_{r,t} r^us^v C_p^m$ with $2n/p$ 
left coset representatives $r^us^v$ where $u=0\ldots n/p-1$ and $v=0,1$.

The second is dihedral $D_{p,\nu}^{n}$ of order $p$ where $p$ is a divisor
of $n$ and $\nu=0\ldots n/p-1$.
There are $2p$ elements of $D_{p,\nu}^{n}$ given by the elements
$r^{qn/p+u\nu}S^t$ of the group $D_n$ for $q=0\ldots p-1$ and $u=0,1$.
The left coset decomposition of the group $D_n$ is given by 
$D_n=\cup_{r} r^v D_{p,\nu}^{n}$ with $n/p$ left coset representatives
$r^v$ where $v=0\ldots n/p-1$.

The projectively inequivalent irreducible 
projective unitary representations of $D_n$ are most easily obtained
from the inequivalent irreducible unitary representations of the 
double dihedral group $D'_n$
with $4n$ group elements  $r^m s^t e^u$ where 
$m=0\ldots n-1$, $t=0,1$, and $u=0,1$.
The generators of the double dihedral group satisfy $r'^n=s'^2=e'$ with 
$e'^2=\mathbf{1}$ where $\mathbf{1}$ is the identity element.
For the generators $r'$, $s'$, $e'$, 
the inequivalent irreducible unitary representations are given by
Table \ref{chap:symmetry:tab:ddn_reps} with the homomorphism relation 
$\phi^{D'_n}_{Rep}(r'^ms'^te'^u)=
\phi^{D'_n}_{Rep}(r')^m\phi^{D'_n}_{Rep}(s')^t\phi^{D'_n}_{Rep}(e')^u$
specifying the representation for the entire group.
\begin{table}
\begin{center}
\begin{tabular}{l|l|lll}
&Rep.&$r'$&$s'$&$e'$\\
\hline
$n$ odd&$A_0$&$+1$&$+1$&$+1$\\
&$A_1$&$+1$&$-1$&$+1$\\
&$B_0$&$-1$&$+i$&$-1$\\
&$B_1$&$-1$&$-i$&$-1$\\
&$E_\mu$&$\exp(\pi i\mu\sigma^u/n)$&$i^\mu\sigma^z$&$(-1)^\mu\sigma^0$\\
\hline
$n$ even&$A_0$&$+1$&$+1$&$+1$\\
&$A_1$&$+1$&$-1$&$+1$\\
&$B_0$&$-1$&$+1$&$+1$\\
&$B_1$&$-1$&$-1$&$+1$\\
&$E_\mu$&$\exp(\pi i\mu\sigma^u/n)$&$i^\mu\sigma^z$&$(-1)^\mu\sigma^0$
\end{tabular}
\end{center}
\caption[Inequivalent irreducible unitary representations of the double dihedral
group $D'_n$]{The 
inequivalent irreducible unitary representations $\phi^{D'_n}_{Rep}$ 
for the generators $r'$, $s'$, $e'$ of the double dihedral group $D'_n$.
Here $\mu=1\ldots n-1$, $\sigma^i$ are the Pauli matrices, and $\sigma^0$
is the identity matrix.}
\label{chap:symmetry:tab:ddn_reps}
\end{table}

When $n$ is odd, the projectively inequivalent irreducible projective
unitary representations of $D_n$
are projectively equivalent to the complex unitary representations of $D_n$
with trivial factor system.
When $n$ is even, it is convenient to introduce the function $f$
which embeds the dihedral group $D_n$ into the double dihedral group
$D'_n$ via $f(R^mS^s)=R'^mS'^s$.  The projectively inequivalent irreducible
projective unitary representations of $D_n$
are projectively equivalent to $\phi^{D_2}(g)=\phi^{D'_2}(f(g))$ where
$g$ is an element in $D_n$ and the factor system is
$\lambda(g_1,g_2)=\phi^{D'_2}(f(g_1g_2)^{-1}f(g_1)f(g_2))$.
This factor system is projectively equivalent to the trivial factor system 
when $\phi^{D'_n}(E')=+1$.  It is a non-trivial factor system for 
$\phi^{D'_n}(E')=-1$ which occurs for the 
$E_\mu$ irreducible unitary representations of $D'_n$
with $n$ even and $\mu$ odd.

\section{\label{chap:symmetry:app:sg}Representation theory approach to spin groups}
Here we compare the implicit classification of spin groups
presented in Litvin and Opechowski \cite{litvin-74} and the constructive
approach in Section \ref{chap:symmetry:sec:sg} using the representation
theory of space groups.

Litvin and Opechowski use a result classifying subgroups of direct
product groups originally due to Zamorzaev \cite{zamorzaev-68}.
Consider the direct product $\mathbf{B}\otimes \mathbf{F}$ of groups 
$\mathbf{B}$, $\mathbf{F}$.
An element of $\mathbf{B}\otimes \mathbf{F}$
is given by $(B,F)$ and the identity,
product, and inverse are given by $(\mathbf{1}_B,\mathbf{1}_F)$,
$(B',F')(B,F)=(B'B,F'F)$, $(B,F)^{-1}=(B^{-1},F^{-1})$, where 
$\mathbf{1}_{B,F}$ is the identity element in $\mathbf{B}$, $\mathbf{F}$.

Denote a subgroup of $\mathbf{B}\otimes \mathbf{F}$ by X.
For all elements of X consisting of elements of the form $(B,F)$,
drop $F$ to obtain $\mathcal{B}$, a subgroup of $\mathbf{B}$.
For all elements of X consisting of elements of the form $(B,F)$,
drop $B$ to obtain $\mathcal{F}$, a subgroup of $\mathbf{F}$.
For all elements of X consisting of elements of the form $(B,\mathbf{1}_F)$,
drop $\mathbf{1}_F$ to obtain a normal subgroup $b$ of $\mathcal{B}$.
For all elements of X consisting of elements of the form $(\mathbf{1}_B,F)$,
drop $\mathbf{1}_B$ to obtain a normal subgroup $f$ of $\mathcal{F}$.
Litvin and Opechowski call $X$ in the family of $\mathcal{B}$ and
$\mathcal{F}$.  The result of Zamorzaev states that the quotient
groups $\mathcal{B}/b$ and $\mathcal{F}/f$ are isomorphic.

Subgroups $X$ of $\mathbf{B}\otimes \mathbf{F}$ are thus classified by 
a normal subgroup $b$ of $\mathcal{B}$ the latter of 
which is a subgroup of $\mathbf{B}$,
a normal subgroup $f$ of $\mathcal{F}$ the latter of 
which is a subgroup of $\mathbf{F}$,
and an isomorphism $\psi$ from $\mathcal{B}/b$ from $\mathcal{F}/f$.

The connection between the Litvin and Opechowski approach and
the representation theory approach is given by the first
isomorphism theorem
\cite{artin-91}.
For a homomorphism $\varphi$ from group $G$ to group $H$,
the first isomorphism theorem states that 
(1) $\text{ker}(\varphi)$ is a normal subgroup of $G$,
(2) $\text{im}(\varphi)$ is a subgroup of $H$, and 
(3) $\text{im}(\varphi)$ is isomorphic to the quotient group 
$G/\text{ker}(\phi$).

Spin groups are subgroups of the direct product group $E(2)\otimes O(3)$
with $E(2)$ the two-dimensional Euclidean group of 
real-space operations and
$O(3)$ the three-dimensional orthogonal group of 
spin-space operations.
Recall that within the representation theory approach, spin groups
are given by a choice of space group $SG$ with elements $(M,t)$, 
a choice of a three-dimensional orthogonal
representation $\phi$, and $N$ is a group that satisfies
$\phi(M,t)^{-1}N\phi(M,t)=N$.

Let us take
$b=\text{ker}(\phi)$, $\mathcal{B}=SG$, $\mathbf{B}=E(2)$, and
$f=N$, $\mathcal{F}=\text{im}(\phi)N=N\text{im}(\phi)$, $\mathbf{F}=O(3)$.
From (1) of the first isomorphism theorem, $b$ is a normal subgroup
of $\mathcal{B}$ and we already know that $SG$ is a subgroup of $E(2)$.
By construction $f$ is a normal subgroup of $\mathcal{F}$ the latter of
which is a
subgroup of $\mathbf{F}$.
The quotient group $\mathcal{F}/f$ is the image
$\text{im}(\phi)$ while $\mathcal{B}/b$ is the kernel
$\text{ker}(\phi$).  From (3) of the first isomorphism theorem,
we see that $\mathcal{B}/b$ and $\mathcal{F}/f$ are isomorphic.

Thus we see that given $SG$, $N$, and $\phi$ within the representation
theory approach, we can construct $b$, $\mathcal{B}$, $f$, $\mathcal{F}$
within the Litvin-Opechowski approach.
If instead we are given $b$, $\mathcal{B}$, $f$, $\mathcal{F}$, we can
again use the first isomorphism theorem to construct $SG$, $N$, and $\phi$.

\section{\label{chap:symmetry:app:space_group}Unitary representations and anti-unitary co-representations of space groups}
In this appendix, we outline the construction of the unitary representations
and anti-unitary co-representations of space groups.  We will use
the results of Appendix \ref{chap:symmetry:subsec:little} on small
representations and little groups
and representation theory as well as the results of Appendix 
\ref{chap:symmetry:app:pg} on point groups in two dimensions.

Recall from Section \ref{chap:symmetry:sec:sg} that an element 
$(M,t)$ of a two-dimensional space group $SG$
consists of a $2\times 2$ orthogonal matrix $M$ describing
rotations/reflections and a two-component vector $t$ describing
translations.
It acts on a point $x$ via
\begin{align}
x_\mu\rightarrow M_{\mu\nu}x_\nu+t_\mu
\end{align}
and the product satisfies
\begin{align}
(M',t')(M,t)=(M'M,M't+t')
\end{align}
where $M$ describes the action of the point group $PG$ and $t$ the action
of the translations $T$.  In particular, this implies
\begin{align}
\nonumber
(M,t)^{-1}&=(M^{-1},-M^{-1}t),\\
(M',t')^{-1}(M,t)(M',t')&=(M'^{-1}MM',M'^{-1}(Mt'+t-t'))
\end{align}
for the inverse element and conjugate action of the element $(M,t)$ by
the element $(M',t')$, respectively.

The translation subgroup $T$ consists of elements $(\mathbf{1},t)$
with $\mathbf{1}$ the identity matrix.  It is an abelian group with 
generators $T_1=(\mathbf{1},t_1)$, $T_2=(\mathbf{1},t_2)$.  
From the above, conjugation of 
$(\mathbf{1},t)$ by $(M',t')$ yields $(\mathbf{1},M'^{-1}t)$.  This implies
$T$ is a normal subgroup of $SG$ and the quotient group $SG/T$ is called
the point group $PG$.  It is either a cyclic or dihedral group of order 
$n=1,2,3,4,6$ as described in Appendix \ref{chap:symmetry:app:pg}.

Since $T$ is a normal and abelian subgroup
of $SG$, we will use the results of Appendix \ref{chap:symmetry:subsec:little}
to construct the inequivalent irreducible unitary representations.
The inequivalent irreducible representations
of $T$ are labeled by a wavevector $k=\gamma k_1+\delta k_2$ where $k_i$
are basis vectors for the reciprocal lattice satisfying
$k_i\cdot t_i=\delta_{ij}$ with $\cdot$ the dot product.  This representation is
given by
\begin{align}
\phi^T_{k}(T_1^cT_2^d)=\exp[-2\pi i k\cdot(ct_1+dt_2)]=\exp[-2\pi i(\gamma c+\delta d)]
\end{align}
where $k$ is restricted to the first Brioullin zone.

The conjugate action is given by 
$\phi^T_{k}((M,t)^{-1}T_1^cT_2^d(M,t))=\phi^T_{Mk}(T_1^cT_2^d)$ from which we can 
see that it is equivalent to the rotation/reflection $M$ acting directly
on the wavevector $k$.  
Under the equivalence relation defined by this conjugate action,
$k$ and $Mk$ are in the same class.
These classes divide the Brioullin zone into $|PG|$ regions with $|PG|$ the
order of the point group.  We then choose
one $k$ as a representative for each class.

For each of these $k$, 
consider the little group $SG_{k}$ given by the subgroup of $SG$
with elements $(M,t)$ that leave $\phi^T_{k}$ fixed under the conjugate action.
Since the conjugate action takes $\phi^T_{k}$ to $\phi^T_{Mk}$, this implies
that $(M,t)$ is in the little group  $SG_{k}$ if $Mk$ and $k$ differ by
a reciprocal lattice vector.

The quotient group $SG_{k}/T$ is a subgroup of the quotient group $SG/T$.
Since the latter is the point group $SG/T=PG$, we will refer to the former
as the wavevector point group $SG_k/T=PG_k$.
In Appendix \ref{chap:symmetry:app:pg},
we list the two-dimensional point groups $PG$ and their possible subgroups.
From \ref{chap:symmetry:tab:space_group}, we list the group elements $(M,t)$
for the point group generators $R$, $S$.  This allows us to obtain the 
left coset representatives $r_i$ for the left coset decomposition
$SG=\cup_i r_i SG_k/T$, the multiplication law $r_1 r_2=r_3(r_1,r_2)$ 
for the quotient
group $SG_k/T$, and the multiplication law $r_1 r_2=r_3(r_1,r_2)h_3(r_1,r_2)$ 
for the group
$SG$ where $h_3(r_1,r_2)$ is an element of the translation group $T$.  

This then gives the factor system
$\phi_k^T(h_3(r_1,r_2))$.  We list
the possible projective representations which $\phi^T_k(h_3(r_1,r_2))$ is 
projectively
equivalent to in Appendix \ref{chap:symmetry:app:pg}.  The
irreducible projective unitary representations $\psi^{PG_k}_{Rep}$
that arise give an irreducible unitary 
representation of $SG_k$.  The induced representation 
$\psi^{PG_k\uparrow SG}_{Rep}$ is an irreducible unitary representation of
$SG$.  Choosing
one $k$ as a representative for the equivalence classes defined by the
relation $k\sim Mk$ gives all of the inequivalent 
irreducible unitary representations of $SG$.

For a given space group $SG$, it is useful to consider two types of 
space groups derived from $SG$ in the construction of 
anti-unitary co-representations: grey space groups $SG^{Grey}$ and 
black-white space groups $SG^{BW}$.
We first introduce the element $\tau$ that commutes with all elements
of $SG$ and satisfies $\tau^{2}=\mathbf{1}$ with $\mathbf{1}$ the identity
element.  A physical interpretation for $\tau$ is as the time-reversal
operator.

A grey space group is given by the left coset decomposition
$SG^{Grey}=SG\cup \tau SG$.  It has
double the number of elements of the original space group $SG$ the latter
of which is a halving subgroup of $SG^{Grey}$.
We have already discussed how the inequivalent irreducible unitary
representations of a space group $SG$ are constructed.
For a grey space group $SG^{Grey}=SG\cup \tau SG$, we can use the
results of Appendix \ref{chap:symmetry:app:au} with the group $G=SG^{Grey}$
and halving subgroup $H=SG$ to then construct the inequivalent irreducible
anti-unitary co-representations.

A black-white space group is given by the left coset decomposition
$SG^{BW}=SG^{1/2}\cup \tau z SG^{1/2}$.
It has the same number of elements as the original space group $SG$.
Here $SG$ itself has a halving subgroup $SG^{1/2}$ 
and left coset decomposition $SG=SG^{1/2}\cup z SG^{1/2}$ where
$z$ is the left coset representative.  Given the inequivalent
irreducible unitary representations of the halving space group
$SG^{1/2}$, we can again use Appendix \ref{chap:symmetry:app:au}
to construct the inequivalent irreducible anti-unitary co-representations.

For each space group $SG$, we see there is only one grey space group $SG^{Grey}$.
However, there can be multiple inequivalent  halving subgroups
$SG^{1/2}$ for $SG$ and thus multiple black-white space groups
$SG^{BW}$ for $SG$.  Here two halving subgroups $SG^{1/2}$ and
${SG^{1/2}}'$ of $SG$ are equivalent if they
are related conjugation by a fixed element $(M,t)$ of the larger $E(2)$
Euclidean group ${SG^{1/2}}'=(M,t)^{-1}SG^{1/2}(M,t)$.
Tables of inequivalent halving space groups for each space group $SG$ are
given in Ref. \cite{litvin-08}.

For black-white space groups, we see that each element
$(M,t)$ of the space group $SG$ is associated with either the element
$(M,t)$ or $\tau(M,t)$ (but not both) 
in the black-white space group $SG^{BW}$.  Here the nomenclature
of black-white space group becomes clear since for $(M,t)$ in $SG$
we can associate the color white if it corresponds to
$(M,t)$ in $SG^{BW}$ and black if it corresponds to $\tau(M,t)$
(or vice versa).  For grey space groups, we see that each element
$(M,t)$ of the space group $SG$ is associated with both of the 
elements $(M,t)$ and $\tau(M,t)$ in the grey space group $SG^{Grey}$.
Using the same nomenclature, each $(M,t)$ in $SG$ is black and white
and associated with the color grey.

There is one difficulty in construction of the anti-unitary co-representations
of a grey $SG^{Grey}$, or black-white space group $SG^{BW}$ 
from the unitary representations of the appropriate
halving space group $SG$.  This lies in the calculation of the unitary matrix $Z$
since one must be careful in defining the sum over the halving
space group $SG$ which is infinite.  Here it is useful to use the left 
coset decomposition of $SG$ by the translation subgroup $T$
given by $SG=\cup_i r_i SG/T$ where $r_i$ are left coset 
representatives of the quotient group $SG/T$ which is given by the point group.  
This allows us to write
$\sum_{SG}=\sum_{r_i}\sum_{T}$.  The summation over $r_i$ corresponds to
a summation over the point group which is finite and well-defined.
The summation over the translation group $T$ corresponds to a discrete
Fourier transform.  Although it is formally an infinite sum, it physically
corresponds to projection of the summand
onto the zero wavevector component which is well-defined.

\section{\label{chap:symmetry:app:example}Spin group for the minimal energy spin texture}

Here we present the construction of the spin group for the minimal energy
spin texture.  This particular spin group is constructed from an anti-unitary
co-representation of a black-white space group.
It offers a illustration of the construction of
irreducible unitary representations and
anti-unitary co-representations of space groups and 
their use in the construction of spin groups.

The space group is given by $p2mg$ with the normal subgroup of translations
given by a rectangular Bravais lattice $T_{Rect}$ with generators given by
\begin{align}
T_1&=(\mathbf{1},[a,0]),& T_2&=(\mathbf{1},[0,b])
\end{align}
where $a$, $b$  are the lattice constants and $\mathbf{1}$ is the $2\times 2$
identity matrix.  The point group given by
the quotient group $p2mg/T_{Rect}$ is the dihedral group $D_2$ of order $n=2$.
This space group is non-symmorphic with generators for rotations $R$ and
reflections $S$ given by
\begin{align}
R&=\left(-\mathbf{1},[0,0]\right),&S&=\left(-\sigma^z,[a/2,0]\right)
\end{align}
where $\sigma^z$ is a Pauli matrix and notice that $S$ has an associated non-trivial
translation.

One of the halving space groups for $p2mg$ is given by the $p2gg$ space group.
It also has a rectangular Bravais lattice $T^{1/2}_{Rect}$ with generators
given by 
\begin{align}
T^{1/2}_1&=(\mathbf{1},[a,0]),& T^{1/2}_2&=(\mathbf{1},[0,2b])
\end{align}
where the lattice constant for the $T^{1/2}_2$ element of the halving space
group $p2gg$ is twice that of $T_2$ for the space group $p2mg$.
Notice in particular that the element $T_2$ of the space group $p2mg$ is
not an element of the halving space group $p2gg$.
The point group given by the quotient group $p2gg/T^{1/2}_{Rect}$
is also the dihedral group $D_2$ of order $n=2$.  This halving space group
is also non-symmorphic with generators for rotations $R^{1/2}$ and reflections
$S^{1/2}$ given by
\begin{align}
R^{1/2}&=\left(-\mathbf{1},[0,0]\right),&S^{1/2}&=\left(-\sigma^z,[a/2,b]\right)
\end{align}
where $\sigma^z$ is a Pauli matrix and notice that $S$ has an associated non-trivial
translation.
The left coset decomposition of the space group $p2mg$ by the halving space group 
$p2gg$ is given by $p2mg=p2gg\cup T_2 p2gg$.  Here the left coset representative
is given by $T_2$.  The corresponding black-white space group is $p(2b)m'g'$
in the notation of Ref. \cite{litvin-08}.

We now turn to the construction of one of the inequivalent irreducible unitary
representations of the halving space group $p2gg$ using the procedure
described in Appendix \ref{chap:symmetry:subsec:little} and
\ref{chap:symmetry:app:space_group}.
The wavevector specifying the irreducible unitary representation of the translation
subgroup $T^{1/2}_{Rect}$ of the halving space group $p2gg$
for the minimal energy spin group
is given by the wavevector $k=k_1/2$ where $k_2=(2\pi/a,0)$ is a reciprocal
lattice vector.  Explicitly, the representation is given by
\begin{align}
\phi^{T^{1/2}_{Rect}}_{k_1/2}((T^{1/2}_1)^c(T^{1/2}_2)^d)=\exp[-\pi i c]
\end{align}
where a general element $t=(T^{1/2}_1)^c(T^{1/2}_2)^d$
of the translation group $T^{1/2}$ is
expressed as $c$, $d$ powers of the generators $T^{1/2}_1$, 
$T^{1/2}_1$.

Conjugation by the generators
$R^{1/2}$, $S^{1/2}$ of the $D_2$ point group 
for the halving space group $p2gg$
leaves $\phi^{T^{1/2}_{Rect}}_{k_2/2}$ fixed.  Since conjugation by $T^{1/2}_1$ 
and $T^{1/2}_2$ also leaves $\phi^{T^{1/2}_{Rect}}_{k_1/2}$ fixed, we see that
the little group $p2gg_{k_1/2}$ for the $k=k_1/2$ representation of the
translation subgroup $T^{1/2}_{Rect}$ of the halving subgroup $p2gg$
is given by $p2gg$ itself.

The quotient group $p2gg_{k_1/2}/T^{1/2}_{Rect}=D_2$ of the little group 
by the translation subgroup is the $D_2$ point group.  Consider the element 
$R^{1/2}S^{1/2}=\left(+\sigma^z,[a/2,b]\right)$.  For the quotient group
$p2gg_{k_1/2}/T^{1/2}_{Rect}=D_2$ we see the multiplication law is 
$R^{1/2}S^{1/2}R^{1/2}S^{1/2}=\mathbf{1}$.  In $p2gg_{k_1/2}$ itself, the 
multiplication law
is $R^{1/2}S^{1/2}R^{1/2}S^{1/2}=T_1^{1/2}$.  Since 
$\phi^{T^{1/2}_1}_{k_1/2}(T_2^{1/2})=-1$ is non-trivial, we see that
we require one of the projectively 
inequivalent irreducible projective unitary representations of 
$p2gg/T^{1/2}_{Rect}=D_2$ with
non-trivial factor system.  

From Appendix \ref{chap:symmetry:app:space_group},
we see there is only one such projective unitary representation of $D_2$
given by the $E_1$ unitary representation of $D'_2$
in Table \ref{chap:symmetry:tab:ddn_reps}.
Labeling this projective representation as
$\psi^{p2gg_{k_1/2}/T^{1/2}_{Rect}}_{E_1}$,
we see it gives one of the inequivalent
irreducible unitary representations of the little group $p2gg_{k_1/2}$.
Since $p2gg_{k/1/2}=p2gg$ is the halving subgroup $p2gg$ itself, the
induced representation $\psi^{p2gg_{k_1/2}/T^{1/2}_{Rect}\uparrow p2gg}_{E_1}$
is simply $\psi^{p2gg_{k_1/2}/T^{1/2}_{Rect}}_{E_1}$.  Thus we obtain
one of the inequivalent
irreducible unitary representations of the halving subgroup $p2gg$.
Labeling this representation as $\phi^{p2gg}_{k_1/2,E_1}$, we find for the
generators
\begin{align}
\nonumber
\phi^{p2gg}_{k_1/2,E_1}(T^{1/2}_1)&=-\sigma^0,&
\phi^{p2gg}_{k_1/2,E_1}(T^{1/2}_2)&=+\sigma^0,\\
\phi^{p2gg}_{k_1/2,E_1}(R^{1/2})&=\sigma^y,&
\phi^{p2gg}_{k_1/2,E_1}(S^{1/2})&=\sigma^z
\end{align}
where $\mathbf{1}$ is the $2\times 2$ identity matrix and $\sigma$
are the Pauli matrices.

Using this irreducible unitary representation of the halving subgroup $p2gg$,
We now turn to the construction of one of the inequivalent irreducible anti-unitary
representations of the space group $p2gg$ using the procedure
described in Appendix \ref{chap:symmetry:app:au} and
\ref{chap:symmetry:app:space_group}. 
The left coset decomposition of $p2mg=p2gg\cup T_2p2gg$ has left
coset representative $T_2$.  The conjugate action of $T_2$ is given by
\begin{align}
\nonumber
T_2^{-1}T^{1/2}_1T_2&=T^{1/2}_1,&
T_2^{-1}T^{1/2}_2T_2&=T^{1/2}_2,\\
T_2^{-1}R^{1/2}T_2&=R^{1/2}T^{1/2}_2,&
T_2^{-1}S^{1/2}T_2&=S^{1/2},&
\end{align}
on the generators of the halving subgroup $p2gg$.
We can check that 
\begin{align}
\phi^{p2gg}_{k_1/2,E_1}(T_2^{-1}hT_2^{-1})^*=\sigma^z\phi^{p2gg}_{k_1/2,E_1}(h)\sigma^z
\end{align}
for each of the elements $h$ of the halving subgroup $p2gg$.  This implies that 
the the unitary matrix $Z=\sigma^z$ with $ZZ^*=+1$ and the
resulting anti-unitary co-representation is of type (2b).
Labeling this anti-unitary co-representation as $\phi^{p2mg}_{k_1/2,E_1,AU}$,
we find for the generators
\begin{align}
\nonumber
\phi^{p2mg}_{k_1/2,E_1,AU}(T_1)&=-\sigma^0,&
\phi^{p2mg}_{k_1/2,E_1,AU}(T_2)&=+\sigma^0\Theta,\\
\phi^{p2mg}_{k_1/2,E_1,AU}(R)&=\sigma^y,&
\phi^{p2mg}_{k_1/2,E_1,AU}(S)&=\sigma^z\Theta
\end{align}
where $\Theta$ is the complex conjugation operator.

Using the results of Appendix \ref{chap:symmetry:app:orth}, we can then calculate
the corresponding real orthogonal representation labelled as 
$\phi^{p2gg}_{k_1/2,E_1,Orth}$.  We find for the generators
\begin{align}
\nonumber
\phi^{p2mg}_{k_1/2,E_1,Orth}(T_1)&=\text{Diag}[+++]
,\\
\nonumber
\phi^{p2mg}_{k_1/2,E_1,Orth}(T_2)&=\text{Diag}[-++]
,\\
\nonumber
\phi^{p2mg}_{k_1/2,E_1,Orth}(R)&=\text{Diag}[-+-]
,\\
\phi^{p2mg}_{k_1/2,E_1,Orth}(S)&=\text{Diag}[+-+]
\end{align}
where $\text{Diag}[s_1 s_2 s_3]$ denotes the $3\times 3$ diagonal
matrix with entries $s_i$ on the diagonal.
The corresponding spin group is associated with the minimal
energy spin texture and is also shown in Table \ref{chap:symmetry:tab:allowed}.

\section{\label{chap:symmetry:app:table}Construction of compatbile spin groups}

\begin{figure}
\begin{center}
\includegraphics[width=3in]{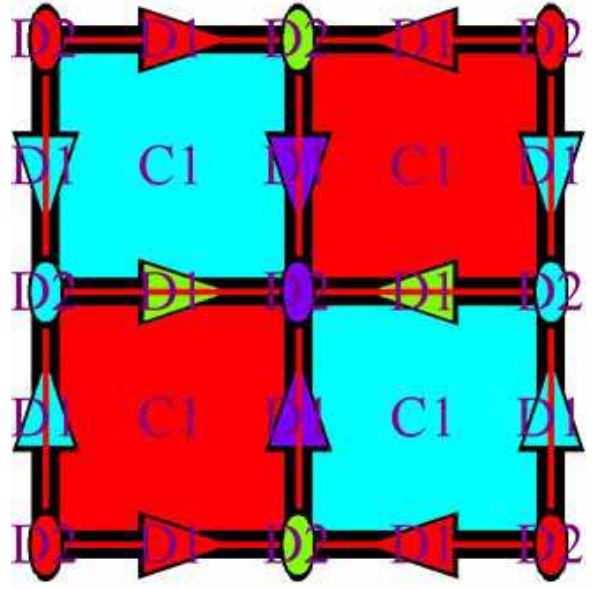}
\end{center}
\caption[Rectangular lattoce Brioullin zone]{
Brioullin zone for a space group $SG$ with rectangular Bravais lattice 
$T_{Rect}$ and $SG/T_{Rect}=D_2$ point group.
Each wavevector $k$ in the Brioullin zone describes an inequivalent
and irreducible representation $\phi_k$ of the translation group $T_{Rect}$.
The little group $SG_k$ consists of elements of $SG$ that leave $\phi_k$
fixed under conjugation.  The wavevector point group
$SG_k/T_{Rect}$ is the trivial group $C_1$ for a generic wavevector in
the Brioullin zone, $D_1$ on high symmetry lines, and $D_2$ on high symmetry
points.
}
\label{chap:symmetry:fig:bz}
\end{figure}

In this section, we discuss the construction of the list compatible
spin groups in Table \ref{chap:symmetry:tab:allowed}.  In 
Section \ref{chap:symmetry:sec:sg} of the main text, we have already
argued that the constraints in Section \ref{chap:symmetry:sec:constraints}
allow us to consider spin groups with real space operations given by
space groups with a $T_{Rect}$ rectangular Bravais lattice and $D_2$ point group.
This implies that the space group $SG$ is either $p2mm$, $p2mg$, or $p2gg$.
In addition, the subgroup of global spin space operations $N$ has to be the trivial
group.
Here we discuss how
to find all of the unitary representations and anti-unitary co-representations
that give rise to real orthogonal representations describing spin space
operations for spin groups.

Let us first consider spin group arising from unitary representations
of space groups.
From Appendix \ref{chap:symmetry:app:space_group}, for each space group $SG$
we first consider all wavevectors $k$ which give rise to
irreducible and inequivalent representations $\phi_k$ of $T_{Rect}$.
This is given by the first Brioullin zone 
is shown for $T_{Rect}$ in Fig. \ref{chap:symmetry:fig:bz}.

Choosing one wavevector $k$ out of each set of wavevectors related
by a point group operation,
we then construct the
wavevector point group $PG_k=SG_k/T_{Rect}$ given by
the quotient group of the little group $SG_k$ leaving $\phi_k$ 
fixed under conjugation by $T_{Rect}$.
We then find all of the projectively inequivalent projective representations
$\psi^{PG_k}_{Rep}$ of $PG_k=SG_k/T_{Rect}$ and then construct
the induced representation $\psi^{PG_k\uparrow SG}_{Rep}$.  This gives
all of the inequivalent and irreducible unitary representations of $SG$.

Notice as $k$ varies continuously throughout
the first Brioullin zone, $\phi_k$ varies continuously.  However,
$PG_k$ changes discontinuously from the trivial group for a generic
point to $D_1$ on high symmetry lines and $D_2$ on high symmetry points.
This means that although individual matrix elements of the 
unitary representations of $SG$ depend continuously on $k$, 
the underlying structure of the unitary representation such as locations
of non-zero matrix elements and dimensionality only change discontinuously
at high symmetry lines and points.  This makes it possible enumerate all of
inequivalent and irreducible unitary representations of $SG$ by treating
all of the generic wavevectors $k$ together and all of the wavevectors $k$ on
each of the high symmetry lines together.

It can be shown that dimensionality of the unitary representation of $SG$ at
a generic wavevector $k$ is given by the order of the point group $PG$.
For $D_2$, the order or number of elements is four which implies that we cannot
use it to construct a three-dimensional real orthogonal representation.
At high symmetry points and lines, the dimensionality of the unitary
representation of $SG$ can be smaller and if it is equal to two, 
it gives rise to a three-dimensional real orthogonal representation 
which can be used to construct a spin group.  Furthermore, it can also
be shown that tuning $k$ on high symmetry lines tunes an incommensurate 
spin modulation
between unit cells living on top of short length scale modulations within
each unit cell.  On physical grounds we expect the shorter length scale
modulations to capture most of the gains in dipolar interaction energy.
Alternatively, a specific value of $k$ on high symmetry lines
is selected based on energetics and we expect that minima occur at the
boundaries corresponding to the high symmetry points.  From these above
arguments, we focus solely on the two-dimensional 
unitary representations arising from high symmetry points.

For each of these two-dimensional unitary representations, 
we then consider the actions of the corresponding spin groups on the magnetization
and skyrmion charge in real space and momentum space as described
in Section \ref{chap:symmetry:sec:sg}.  The spin groups that
do not force the magnetization to vanish at some point but force
the net magnetization and net skyrmion charge to vanish are then
compatible spin groups.  There are only two compatible spin groups
arising from unitary representations of space groups and are the first
ones shown in Table \ref{chap:symmetry:tab:allowed} for the $p2mg$, $p2gg$
space groups.  The corresponding wavevector $k$, wavevector point group $PG_k$
and projective representation $\psi^{PG_k}$ are specified as well.
The projective representations use the notation of Appendix
\ref{chap:symmetry:app:pg}.

Now we consider spin groups arising from anti-unitary co-representations of 
space groups.  From \ref{chap:symmetry:app:space_group}, 
for each space group $SG$ we have to consider
the grey space group $SG^{Grey}$ and each of the black-white space groups
$SG^{BW}$ arising from each of the inequivalent halving subgroups $SG^{1/2}$
of $SG$.
We can rule out the grey space group $SG^{Grey}$ because this gives rise
to a non-trivial global spin symmetry operation.  For each of the 
$SG^{BW}$, we have to first construct the unitary representations
of the halving space groups.  This follows from the same procedure described above
but we have to keep track of both two-dimensional and one-dimensional
unitary representations of $SG^{1/2}$.  This is because it is possible
to construct two-dimensional
anti-unitary co-representations of $SG$ from
one two-dimensional unitary representation of $SG^{1/2}$ or a pair of
one-dimensional unitary representations of $SG^{1/2}$.

Applying the constraints coming from the real space and momentum space
actions of spin groups gives the remaining compatible spin groups 
in Table \ref{chap:symmetry:tab:allowed}.  The name of the black-white
space group $BWSG$ in the notation of Ref. \cite{litvin-08}, the 
halving space group $SG^{1/2}$ are also included.
The corresponding wavevector $k$, wavevector point group $PG_k$
and projective representation $\psi^{PG_k}$ for the halving space group
$SG^{1/2}$ (not the space group $SG$) are specified as well.
The projective representations use the notation of Appendix
\ref{chap:symmetry:app:pg}.

\section{\label{chap:symmetry:app:spherical}Spin texture spherical interpolation}
The low-energy effective theory we consider is defined in terms of a 
three-component unit vector $\hat{n}(x)$ which lives on the sphere
in spin-space.  To numerically calculate the energy, it is necessary
to discretize the spin texture.
In order to accurately describe a smooth spin texture,
it is desirable to interpolate between the discretized values 
before calculating the energy.

In this section, we consider the problem of
spherical interpolation between discrete samples of $\hat{n}(x)$.
We take the spin texture to be in the symmetry class described by a
spin group with basis vectors $t_1$ and $t_2$ and consider
a $N_1\times N_2$ discretization given by
\begin{align}
\hat{n}(u_1,u_2)=\hat{n}(u_1t_1/N_1+u_2t_2/N_2)
\end{align}
where $0\le u_i< N_i$.
Given samples on the corners of a plaquette 
\begin{align}
\nonumber
\begin{bmatrix}
\hat{n}(0.0,1.0)&
\hat{n}(1.0,1.0)\\
\hat{n}(0.0,0.0)&
\hat{n}(1.0,0.0)\\
\end{bmatrix}
\rightarrow\\
\begin{bmatrix}
\hat{n}(0.0,1.0)&
\hat{n}(0.5,1.0)&
\hat{n}(1.0,1.0)\\
\hat{n}(0.0,0.5)&
\hat{n}(0.5,0.5)&
\hat{n}(1.0,0.5)\\
\hat{n}(0.0,0.0)&
\hat{n}(0.5,0.0)&
\hat{n}(1.0,0.0)\\
\end{bmatrix}
\end{align}
we wish to
interpolate samples on the perimeter and interior of the plaquette.

First we consider the problem for the plaquette perimeter.
Consider the points $\hat{n}(0.0,0.0)$, $\hat{n}(1.0,0.0)$,
$\hat{n}(1.0,1.0)$, $\hat{n}(0.0,1.0)$ in counterclockwise order
where for each segment, we need to interpolate between its endpoints.
For example, we need to define $\hat{n}(0.5,0.0)$ on the segment
$\hat{n}(0.0,0.0)\rightarrow\hat{n}(1.0,0.0)$.
Denote the initial and final points on the sphere in 
spin-space as $\hat{n}_i$ and $\hat{n}_f$.
We use geodesics on the sphere consisting of great circles 
in order to describe a trajectory from $\hat{n}_i$ to $\hat{n}_f$
with minimal length.  Explicitly, we take
\begin{align}
\hat{n}(t)&=
\frac{\sin[\gamma (1-t)]}{\sin[\gamma]}\hat{n}_i+
\frac{\sin[\gamma t]}{\sin[\gamma]}\hat{n}_f,&
\cos(\gamma)&=\vec{n}_i\cdot\vec{n}_f
\label{chap:symmetry:eq:geodesic}
\end{align}
from which one can show $\hat{n}(t)\cdot\hat{n}(t)=1$ ensuring
$\hat{n}(t)$ lies on the sphere with $\cdot$ the dot product.
In addition, 
$\hat{n}_i\cdot\hat{n}(t)=\cos[\gamma t]$ and
$\hat{n}_f\cdot\hat{n}(t)=\cos[\gamma (1-t)]$ 
demonstrating the corresponding angles which measure distance on a 
sphere are linear in $t$.

Next we consider the plaquette interior.  
Consider again the points $\hat{n}(0.0,0.0)$, $\hat{n}(1.0,0.0)$,
$\hat{n}(1.0,1.0)$, $\hat{n}(0.0,1.0)$ in counterclockwise order.
By connecting each segment by geodesics as in Eq.
\ref{chap:symmetry:eq:geodesic}, we trace out a region $P$ 
bounded by a closed curve on the sphere with an interior defined
by the right hand rule.  One possible interpolation for the
interior point $\hat{n}(0.5,0.5)$ is the centroid of $P$. 
Since $P$ lives on the sphere, the centroid of the complement
$P^C$ is also
a sensible interpolation for $\hat{n}(0.5,0.5)$.
These two centroids $\vec{m}$, $\vec{m}^C$ are given by
\begin{align}
\vec{m}&=\frac{\int_P dA\ \hat{n}}{\int_P dA},&
\vec{m}^C&=\frac{\int_{P^C} dA\ \hat{n}}{\int_{P^C} dA}
=\frac{-\int_P dA\ \hat{n}}{4\pi-\int_P dA}
\end{align}
where $dA$ is the area element on the sphere and $\hat{n}$ is the normal
on the sphere.  We resolve this ambiguity by selecting the region with
the smallest area from $P$ and $P^C$.  This gives the plaquette
interior point as $\hat{n}(0.5,0.5)=\vec{m}/|\vec{m}|$ if 
$\int_P dA\le2\pi$ and $\hat{S}(0.5,0.5)=\vec{m}^C/|\vec{m}^C|$ otherwise.

The area integral in the dominator can be calculated explicitly for
a region given by a spherical polygon consisting of $M$ points
$\hat{n}_0\ldots\hat{n}_{M-1}$ connected by geodesics of the form
in Eq. \ref{chap:symmetry:eq:geodesic}.
It is given by
\begin{align}
\nonumber
{\int_P dA}&=\sum_{m=0}^{M-1} \theta_m-(M-2)\pi,\\
\tan(\theta_m)&=
\frac{\hat{n}_{m-1}\cdot (\hat{n}_{m}\times\hat{n}_{m+1})}
{\hat{n}_{m-1}\cdot\hat{n}_{m+1}-
(\hat{n}_{m-1}\cdot\hat{n}_{m})
(\hat{n}_{m+1}\cdot\hat{n}_{m})}
\end{align}
where $\theta_n$ is the interior angle defined by the three points
$\hat{n}_{n-1}$, $\hat{n}_{n}$, $\hat{n}_{n+1}$,
indices are taken modulo $M$, $\times$ denotes the cross product,
and $\cdot$ denotes the dot product.
The center of mass integral in the numerator can be calculated via
Stokes theorem
\begin{align}
\nonumber
\int_P dA\ \hat{n}=&
\frac{1}{2}\int dt \hat{n}(t)\times\frac{d\hat{n}(t)}{dt}=\\
&\frac{1}{2}\sum_{m=0}^{M-1} \hat{n}_{m-1}\times\hat{n}_m
\frac{\text{arccos}(\hat{n}_{m-1}\cdot\hat{n}_{m})}
{\sqrt{1-(\hat{n}_{m-1}\cdot\hat{n}_{m})^2}}
\end{align}
where $\hat{n}(t)$ parametrizes the geodesic defining the
boundary of $P$ and indices are taken modulo $N$.

\bibliography{refs}
\bibliographystyle{apsrev}
\end{document}